\documentclass[10pt]{article}

\usepackage{float}
\usepackage[utf8]{inputenc}
\usepackage[T1]{fontenc}
\usepackage{amsmath,amssymb,amsfonts}
\usepackage{graphicx}
\usepackage{hyperref}
\usepackage{url}
\usepackage{algorithm}
\usepackage{algpseudocode}
\usepackage{booktabs}
\usepackage{natbib}
\usepackage{multirow}
\usepackage{setspace}
\usepackage{float}

\usepackage{multicol}

\usepackage{titlesec}

\usepackage[margin=1in]{geometry}

\newcommand{\keywords}[1]{\par\vspace{0.5em}\noindent\textbf{Keywords:} #1\par\vspace{0.5em}}

\begin{document}

\title{Estimating Interpretable Heterogeneous Treatment Effect with Causal Subgroup Discovery in Survival Outcomes }

\author{
Na Bo\footnote{bon@vcu.edu; Department of Biostatistics, Virginia Commonwealth University; Department of Biostatistics and Health Data Science, University of Pittsburgh} \ and Ying Ding\footnote{yingding@pitt.edu; Department of Biostatistics and Health Data Science, University of Pittsburgh}
}

\date{}
\maketitle

\begin{abstract}
Estimating heterogeneous treatment effects (HTE) for survival outcomes has gained increasing attention in precision medicine, as it captures variations in treatment efficacy among patients or subgroups. However, most existing methods conduct post-hoc subgroup identifications rather than simultaneously estimating HTE and identifying causal subgroups. In this paper, we propose an interpretable HTE estimation framework that integrates meta-learners with tree-based methods to estimate the conditional average treatment effect (CATE) for survival outcomes and identify predictive subgroups simultaneously. We evaluated the performance of our method through extensive simulation studies. We also demonstrated its application in a large randomized controlled trial (RCT) for age-related macular degeneration (AMD), a progressive polygenic eye disease, to estimate the HTE of an antioxidant and mineral supplement on time-to-AMD progression and to identify genetically defined subgroups with enhanced treatment effects. Our method offers a direct interpretation of the estimated HTE and provides evidence to support precision healthcare.
\end{abstract}

\section{Introduction}
\label{sec:intro}
Estimating heterogeneous treatment effects (HTE) has gained significant attention in healthcare research, as it captures variations in treatment efficacy between individuals or subgroups exposed to the same treatment. In particular, there is increasing interest in estimating HTE for survival outcomes, which provides a deeper understanding of treatment heterogeneity in disease progression and helps to develop targeted therapeutic strategies tailored to patients with specific characteristics.

In previous studies, subgroup identification methods have been widely explored as an approach to characterizing treatment heterogeneity, yielding subgroups defined by patient characteristics. For example, \cite{ChangePoint_subgroup_JASA2017_Fan} adopted a change-point technique to examine whether a subgroup exists. \cite{AssesVulnerableSubgroup_JASA2023_Guo} focused on pre-defined subgroups and revealed which subgroup was more vulnerable through a sparse logistic regression model, with the aim of obtaining accurate point estimation and valid inference of coefficients. Tree-based methods are also popular as they provide a clearer interpretation of subgroups through a hierarchical tree structure. For example, \cite{tree_subgroup_SIM2023_LuoGuo} developed a tree-based subgroup identification method that accounts for both subgroup size and effect size through a multiple testing procedure when splitting at tree nodes. \cite{subgroup_recrusive_SIM2011_Lipkovich} proposed a recursive partitioning approach with Type I error control that maximizes the difference in treatment effects between two child nodes. \cite{tutorial_subgroup_SIM2016_Lipkovich} also wrote a tutorial summarizing subgroup identification methods into three categories: confirmatory subgroup analysis, post-hoc subgroup evaluation, and subgroup discovery through machine learning. However, subgroup identification methods primarily consider Type I error control for false discovery of subgroups, rather than focusing on estimating HTE.

HTE is assessed by estimating the conditional average treatment effect (CATE), typically defined as the expected difference between potential outcomes in the treatment and control arms, given patient characteristics. Existing methods for CATE estimation in survival outcomes primarily focus on the accurate estimation of CATE and post-hoc subgroup identification. For instance, \cite{CATE_simultaneousCI_Biostatistics2011_Cai} proposed to construct simultaneous confidence intervals for the average treatment effect in subgroups defined by patients sharing the same CATE estimates. \cite{metalearner_surv_JDS,pseudoreg_survival_Bo2024} employed meta-learners with machine learning to estimate CATE for survival data, subsequently identifying beneficial subgroups by selecting thresholds on predicted CATEs. \cite{HTE_mediansurv_ML_Hu2021,clusteredCATEsurv_Hu2022} used Bayesian methods to estimate CATE on the median survival time scale, identifying subgroups by examining posterior average survival effects against covariates. 
\cite{CSF_JRSSb2023} introduced causal survival forests (CSF) to improve the interpretability of CATE estimates through tree structures, although subgroup identification remained challenging due to the complexity of the ensemble trees. 

These methods utilizing machine learning techniques often lack interpretability due to their ``black box'' nature, limiting their interpretations in clinical practice. To address this challenge, there is a growing need for interpretable approaches to CATE estimation. For example, \cite{CRE_havard2020} proposed a causal rule ensemble method to estimate CATE for continuous outcomes. Their two-step approach begins by generating potential subgroups using tree-based methods, where predicted CATEs using existing causal inference methods (e.g., causal forests) are used as outcomes and regressed on covariates. In the second step, a penalized regression is applied, where the predicted CATEs are regressed on the generated subgroups. \cite{wu2023causalrulelearningenhancing} proposed a similar two-step approach, first generating potential subgroups from causal forests and then estimating CATE through D-learner using the generated subgroups as new covariates. 
\cite{RuleEnsemble_KeWan_SIM2023} proposed an interpretable CATE estimation method for continuous outcomes, 
focusing on the main effects of treatment. They first generated subgroups by regressing inverse probability-weighted outcomes on covariates using gradient-boosting trees. Then, they employed an adaptive group lasso to fit the observed outcomes to these subgroups. \cite{RuleEnsemblePrognostic_SMMR2024} extended it to incorporate prognostic effects.

Interpretable CATE estimation in survival outcomes remains under-explored. Only recently, \cite{wan2023survivalcausalruleensemble} extended their previous work \citep{RuleEnsemble_KeWan_SIM2023} using a similar approach based on the Cox proportional hazards (PH) model. In this paper, we propose an interpretable HTE estimation framework to estimate CATE in survival outcomes, which not only predicts CATE but also identifies important subgroups. Our study is motivated by age-related macular degeneration (AMD), a progressive polygenic neurodegenerative eye disease that can cause blindness in older individuals. Randomized clinical trials (RCTs), known as the Age-Related Eye Diseases Study (AREDS and AREDS2), evaluated the efficacy of the AREDS formula (an antioxidant and mineral supplement) in delaying AMD progression, but showed minimal overall effects. \cite{metalearner_surv_JDS} estimated the CATE of the AREDS formula on delaying time-to-AMD progression and identified key Single Nucleotide Polymorphisms (SNP) using the post-hoc algorithm based on predicted CATEs. However, this approach lacks interpretation. In this study, we leverage the interpretability of tree nodes from tree-based algorithms to develop an interpretable framework to estimate CATE in survival outcomes. Instead of using the Cox PH model, which relies on the PH assumption, as in \cite{wan2023survivalcausalruleensemble}, we employ inverse probability censoring weighting (IPCW) to appropriately account for censored observations. Our proposed framework utilizes the idea of ``pseudo-individualized treatment effect'' (pseudo-ITE)\citep{pseudoreg_survival_Bo2024} to address the fundamental causal inference challenge that only one potential outcome is observable for each individual. This framework integrates three meta-learners with different statistical properties to construct pseudo-ITE. 

The paper is organized as follows. In Section \ref{sec:notation}, we introduce the notation, assumptions, and target estimand. In Section \ref{sec:method}, we describe the proposed framework. In Section \ref{sec:simu}, we examine the proposed methods in terms of prediction and subgroup identification accuracy through comprehensive simulations. We applied our proposed method to the AREDS trial to estimate CATE and identify subgroups with differential efficacy in Section \ref{sec:realdat}. Finally, we conclude and discuss limitations and future work in Section \ref{sec:conclusion}.

\section{Problem Set-up}
\label{sec:notation}
The HTE is defined through the CATE function. We follow the Neyman–Rubin counterfactual framework \citep{rubin1974,Neyman1990} to define the CATE for survival outcomes. Consider a study of sample size $n$, which compares two treatment arms, $A_i \in \{0, 1\}$ where $A_i=1$ denotes taking the treatment and $A_i=0$ denotes taking the control for subject $i$, $i \in \mathcal{I}=\{1,2,\cdots,n\}$. $\boldsymbol{X}_i \in \mathbb{R}^p$ is a $p-$dimensional vector of covariates. Let $T$ denote the survival time and $C$ denote the censoring time. $T_i(1)$ and $T_i(0)$ denote potential survival times; $C_i(1)$ and $C_i(0)$ denote potential censoring times; and $\delta_i(1)$ and $\delta_i(0)$ denote potential event indicators. We define $U=\min(T,C)$ as the observed time with the event indicator $\delta$ defined as $\delta=I(T<C)$. The observed data are denoted as $D = \{(U_i, \delta_i), A_i, \boldsymbol{X}_i; i=1, \dots, n \}$. In this paper, we define CATE as the expected difference between $I(T(1)>t^*)$ and $I(T(0)>t^*)$ given covariates:
\begin{eqnarray}
\label{eq:cate_def_survp}
\tau(\boldsymbol{x};t^*)=\mathbb{E}[I(T(1)>t^*)- I(T(0)>t^*)| \boldsymbol{X}=\boldsymbol{x} ],
\end{eqnarray}
where $t^*$ denotes a pre-specified time of interest. 
CATE can also be defined for other survival quantities. For example, on the restricted mean survival time scale, CATE is defined as $\tau(\boldsymbol{x};t^*) = \mathbb{E}[\min(T(1),t^*)-\min(T(0),t^*)|\boldsymbol{X}=\boldsymbol{x}]$, where $t^*$ is the pre-specified restricted time.

In this paper, we focus on CATE under the survival probability scale, defined in Equation (\ref{eq:cate_def_survp}). We assume the following causal analysis assumptions:
\begin{itemize}
\item \textbf{Assumption 1} (Consistency):
 $ T_i=A_iT_i(1)+(1-A_i)T_i(0)$, $C_i=A_iC_i(1)+(1-A_i)C_i(0)$, $\delta_i=A_i\delta_i(1)+(1-A_i)\delta_i(0) $

\item \textbf{Assumption 2} (Unconfoundedness):
$ A_i \!\perp\!\!\!\perp (T_i(0), T_i(1))|\boldsymbol{X}_i$, \\
 $ A_i \!\perp\!\!\!\perp (C_i(0), C_i(1))|\boldsymbol{X}_i $

\item \textbf{Assumption 3} (Population Overlap): 
$
    e(\boldsymbol{x})=P(A_i=1|\boldsymbol{X}_i=\boldsymbol{x}) \in (0,1)
$
\end{itemize}
where $e(\boldsymbol{x})$ is a propensity score for treatment assignment. Given Assumptions 1 to 3, the causal estimand in Equation (\ref{eq:cate_def_survp}) can be identified as
\begin{align}
\label{eq:cate_iden_general}
\tau(\boldsymbol{x};t^*)&=\mathbb{E}[I(T>t^*)|\boldsymbol{X}=\boldsymbol{x}, A=1]-\mathbb{E}[I(T>t^*)|\boldsymbol{X}=\boldsymbol{x}, A=0] \\ \nonumber
&=S_1(t^*|\boldsymbol{x})-S_0(t^*|\boldsymbol{x})
\end{align}
where $S_0(t^*|\boldsymbol{x})$ and $S_1(t^*|\boldsymbol{x})$ are conditional survival probabilities under control and treatment, given covariates. We also assume the standard noninformative censoring given covariates and treatment: $T_i(a) \!\perp\!\!\!\perp C_i(a) |  \{\boldsymbol{X}_i, A_i\}$. Consistent with the assumption in \cite{KennedyDR}, we consider the true CATE function to be a smooth function of $x$, which ensures that it can be well approximated by typical machine learning–based estimators.

\section{Methods}
\label{sec:method}
In this section, we introduce the proposed framework for estimating interpretable HTE in survival outcomes. 

\subsection{Introduction to interpretable machine learning method -- RuleFit}
\label{sec: method_motivation}
In our proposed interpretable HTE estimation framework, we employ an interpretable machine learning algorithm called ``RuleFit'' \citep{RuleFit2008}. Below, we provide an overview of the core idea behind this algorithm.

\begin{figure}
\centering
\includegraphics[width=6.8in]{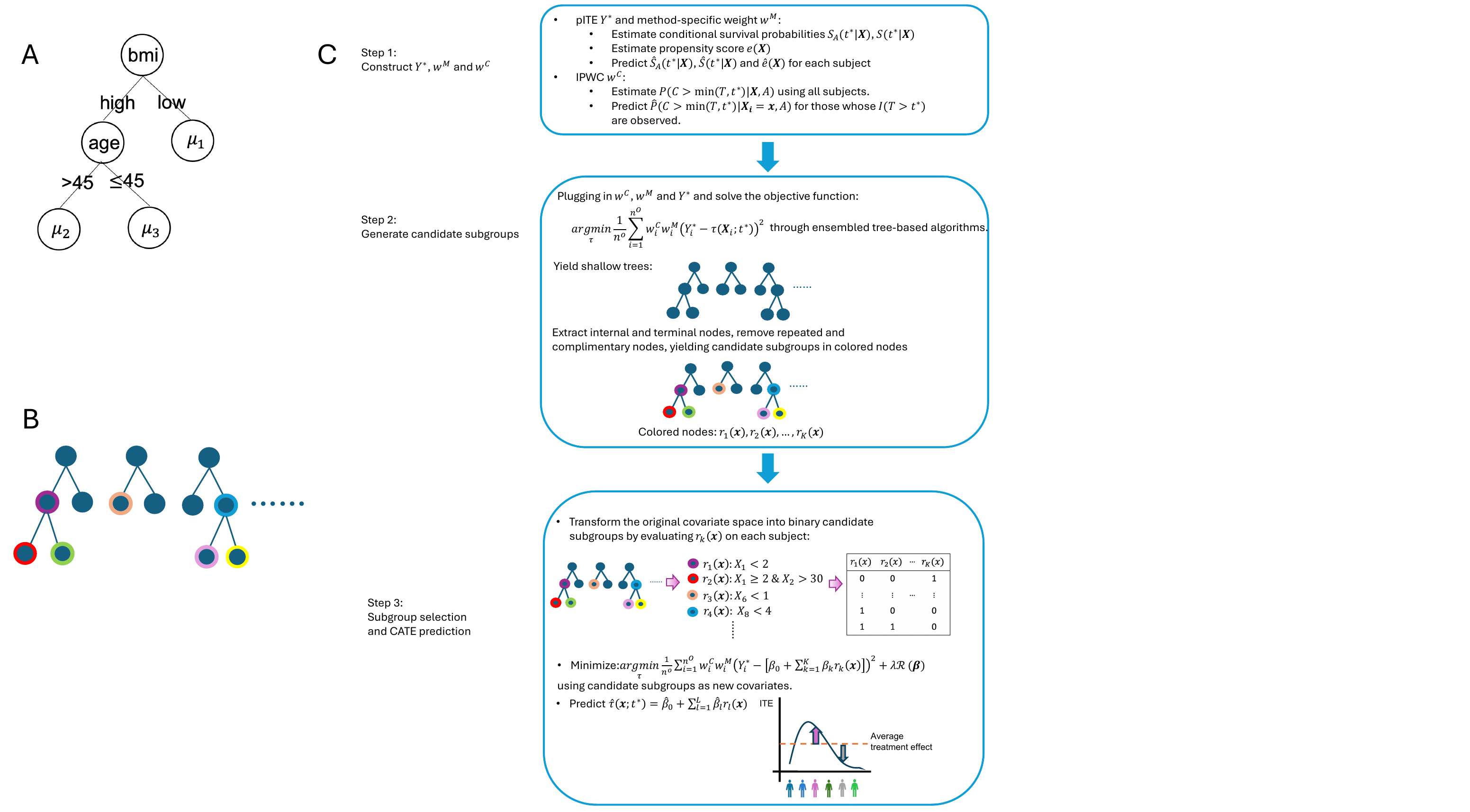}
\caption{Figure A illustrates an example of candidate subgroups generated from a single decision tree. Figure B presents an example of rule extraction from an ensemble of trees. Figure C shows the flowchart of the proposed interpretable meta-learner framework for estimating CATE in survival outcomes.}
\label{fig:example_rule_tree}
\end{figure}

To enhance interpretability and support evidence-based decision-making, the RuleFit approach constructs a prediction model of the form $y=\beta_0+\sum_{k=1}^K \beta_k r_k(\boldsymbol{x})$, where $y$ is a continuous outcome variable, $r_k(\boldsymbol{x}) \in \{0,1\}$, $k=1,2,\cdots, K$, are binary ``rule" terms, and $\beta_k$ are their corresponding coefficients. These rule terms represent simple logical conditions, such as whether a patient meets certain clinical criteria, and are derived from decision trees. Figure \ref{fig:example_rule_tree}A gives an example in which the rule terms are generated from a classification and regression tree (CART) by regressing the outcome $y$ on the covariates. Each internal and terminal node corresponds to a binary condition that can be interpreted as a rule. From the tree of this example, we extract four rule terms. Rule 1: $r_1(\boldsymbol{x})=I(\text{bmi = high})$, Rule 2: $r_2(\boldsymbol{x})=I(\text{bmi = low})$, Rule 3: $r_3(\boldsymbol{x})=I(\text{bmi = high})I(\text{age} > 45)$, and Rule 4: $r_4(\boldsymbol{x})=I(\text{bmi = high})I(\text{age} \leq 45)$. To reduce redundancy, complementary rules are removed. In this case, Rule 2 is excluded because it is complementary to Rule 1. The remaining rules (Rules 1, 3, and 4) define interpretable subgroups, allowing each subject to be evaluated for membership in these subgroups. By generating an ensemble of such trees, the covariate space $X\in \mathbb{R}^p$ is partitioned into binary indicators $r_1(\boldsymbol{x}), r_2(\boldsymbol{x}),\cdots, r_K(\boldsymbol{x})$. These indicators are then treated as new covariates in constructing an interpretable predictive model.

In this paper, we refer to each rule as a candidate subgroup and use these candidate subgroups to construct a prediction model for estimating CATE with survival outcomes. In Section \ref{sec: method_pITE}, we present the general pseudo-ITE framework \citep{pseudoreg_survival_Bo2024} for estimating CATE in the context of survival data. Then, in Section \ref{sec: method_interpretable_ITE}, we introduce our proposed interpretable HTE estimation framework, which applies the RuleFit algorithm within the pseudo-ITE approach.

\subsection{Estimating CATE through a pseudo-ITE-based meta-learner framework}
\label{sec: method_pITE}

As shown in Equation (\ref{eq:cate_def_survp}), if the ITE, defined as $I(T_i(1)>t^*)- I(T_i(0)>t^*)$, were observable for each subject $i$ at a pre-specified time point $t^*$, we could directly regress ITE on covariates to estimate CATE. However, the ITE is unobservable for two main reasons. First, even in the absence of censoring, we can only observe one of the two potential outcomes $I(T(1)>t^*)$ or $I(T(0)>t^*)$, since each patient receives only one treatment. Second, in the presence of censoring, $I(T>t^*)$ cannot be determined for individuals censored before time $t^*$. To address these challenges, \cite{pseudoreg_survival_Bo2024}  proposed to construct a pseudo-ITE (pITE) $Y^*$ based on observed data and estimate the CATE function $\tau(\boldsymbol{x};t^*)$ by minimizing a squared error-type objective function: 
\begin{eqnarray}
\label{loss:squared_error_no_censor}
    \operatorname*{argmin}_{\tau}\frac{1}{n} \sum_{i=1}^{n} w_i^M\left(Y_i^*-\tau(\boldsymbol{X}_i;t^*)\right)^2,
\end{eqnarray}
where $Y_i^*$ is a constructed pITE that replaces the unobservable ITE; $w_i^M$ is a method-specific weight, which will be introduced in the following paragraphs.

To accommodate right-censored data, IPCW, denoted as $w^C= \\ \frac{1}{P(C> \min(T,t^*)|\boldsymbol{X},A)}$, was used to predict the inverse of the probability of not being censored before $\min (T, t^*)$ for those with $I(T>t^*)$ known. Thus, the final objective function to solve for CATE for right-censored data becomes:
\begin{eqnarray}
\label{loss:squared_error_censor}
    \operatorname*{argmin}_{\tau}\frac{1}{n^o} \sum_{i=1}^{n^o} w_i^C w_i^M\left(Y_i^*-\tau(\boldsymbol{X}_i;t^*)\right)^2, 
\end{eqnarray}
where $n^o$ denotes the sample size of complete data whose $I(T>t^*)$ is known.

There are multiple ways to construct $Y^*$, and in this section, we illustrate three doubly-robust approaches to construct pITE.
These three methods, DR-learner, DEA-learner, and R-learner, were originally introduced in \cite{pseudoreg_survival_Bo2024} and are referred to as ``meta-learners.'' Each method constructs pITE with distinct statistical properties. 

\textbf{DR-learner}. \cite{KennedyDR} proposed a Doubly-Robust learner (DR-learner) derived from the efficient influence curve to estimate the average treatment effect. When formulated on the survival probability scale, the pITE can be written as follows:
\begin{eqnarray}
    Y^*_{DR}=\frac{A-\hat{e}(\boldsymbol{X})}{\hat{e}(\boldsymbol{X})\left(1-\hat{e}(\boldsymbol{X})\right)}\left(I(T>t^*)-\hat{S}_A(t^*|\boldsymbol{X})\right)+\hat{S}_1(t^*|\boldsymbol{X})-\hat{S}_0(t^*|\boldsymbol{X}). \nonumber
\end{eqnarray} 
It satisfies $\tau(\boldsymbol{x};t^*)=E[Y^*_{DR}|\boldsymbol{X}=\boldsymbol{x},t^*]$ and does not require a method-specific weight $w^M$ (or equivalently $w^M=1$ for all subjects).

\textbf{DEA-learner}. \cite{DEAlearning2017} proposed the DEA-learner as an extension of the D-learner \citep{Dlearning_TianLu2014}, also known as the modified covariate approach. The original objective function of D-learner solves CATE in a linear parametric form by directly minimizing an objective function, which (on the survival probability scale) can be formulated as follows: 
\begin{eqnarray}
     \frac{1}{n^o} \sum_{i=1}^{n^o} \frac{(2A_i-1) \left(A_i-e\left(\boldsymbol{X}_i\right)\right)}{e\left(\boldsymbol{X}_i\right)\left(1-e\left(\boldsymbol{X}_i\right)\right)}\left(I(T_i>t^*)-\frac{2A_i-1}{2} \boldsymbol{X}_i\gamma\right)^2,\nonumber
\end{eqnarray}
where $\tau\left(\boldsymbol{X}_i; t^*\right)=\boldsymbol{X}_i\gamma$. To relax the restriction in linear parametric form, one can replace $\boldsymbol{X}_i\gamma$ with a general $\tau\left(\boldsymbol{X}_i; t^*\right)$ in the objective function. Since $(2A-1)^2=1$ always holds for both $A=0$ and $A=1$, we can reorganize the objective function as follows:
\begin{eqnarray}
    \frac{1}{n^o}\sum_{i=1}^{n^o} (2A_i-1) \frac{A_i-e\left(\boldsymbol{X}_i\right)}{4e\left(\boldsymbol{X}_i\right)\left(1-e\left(\boldsymbol{X}_i\right)\right)}\left\{2(2A_i-1) I(T_i>t^*)-\tau\left(\boldsymbol{X}_i; t^*\right)\right\}^2, \nonumber
\end{eqnarray}
where $Y_{D}^*=2(2 A-1) I\left(T>t^*\right)$ is the pITE. A regression weight, given by $w^M=(2A-1) \frac{A-e\left(\boldsymbol{X}\right)}{4e\left(\boldsymbol{X}\right)\left(1-e\left(\boldsymbol{X}\right)\right)}$, needs to be applied to regress the pTIE $Y^*_D$ on covariates. 

\cite{DEAlearning2017} modified the D-learner into an efficiency augmentation version (namely, the ``DEA-learner'') by replacing the observed outcome $I(T>t^*)$ with its residual $I(T>t^*)-S(t^*|\boldsymbol{X})$. Thus, $Y_{D E A}^*=\\2\left(2 A-1\right)\left(I\left(T>t^*\right)-\hat{S}(t^*|\boldsymbol{X})\right)$ is the new pITE with the same method-specific weight $w^M=(2A-1) \frac{A-e\left(\boldsymbol{X}\right)}{4e\left(\boldsymbol{X}\right)\left(1-e\left(\boldsymbol{X}\right)\right)}$. 

\textbf{R-learner}. \cite{Rlearner2021} proposed the R-learner using Robinson's decomposition: 
$$ I(T_i>t^*)-E\left[I(T_i>t^*)|\boldsymbol{X}_i=\boldsymbol{x}, t^*\right]=\left(A_i-e\left(\boldsymbol{X}_i\right)\right) \tau\left(\boldsymbol{X}_i;t^*\right)+\varepsilon_i,$$
where $E\left[\varepsilon_i|\boldsymbol{X}_i, A_i\right]=0$. Then, CATE can be solved by minimizing the objective function:
\begin{eqnarray}
    \frac{1}{n^o} \sum_{i=1}^{n^o}\left[\left(I(T_i>t^*)-S\left(t^*|\boldsymbol{X}_i\right)\right)-\left(A_i-e\left(\boldsymbol{X}_i\right)\right) \tau\left(\boldsymbol{X}_i;t^*\right)\right]^2, \nonumber
\end{eqnarray} 
where $S\left(t^*|\boldsymbol{X}\right)$ is the survival function that ignores the treatment groups. We can rewrite the objective function as follows:
\begin{eqnarray}
    \frac{1}{n^o} \sum_{i=1}^{n^o}\left(A_i-e\left(\boldsymbol{X}_i\right)\right)^2\left\{\frac{I(T_i>t^*)-S\left(t^*|\boldsymbol{X}_i\right)}{A_i-e\left(\boldsymbol{X}_i\right)}-\tau\left(\boldsymbol{X}_i;t^*\right)\right\}^2. \nonumber
\end{eqnarray}
Thus, $Y_{RL}^*=\frac{I\left(T>t^*\right)-{S}(t^*|\boldsymbol{X})}{A-{e}(\boldsymbol{X})}$ is the pITE under this approach and $w^M=\left(A-{e}(\boldsymbol{X})\right)^2$ is the method-specific weight.

The propensity score $e(\boldsymbol{X})$ and the mean outcome regression models (i.e., $S_A(t^*|\boldsymbol{X})$ and $S(t^*|\boldsymbol{X})$) in the construction of $Y^*$ and the weights $w^M$, $w^C$ are all treated as nuisance parameters and need to be estimated beforehand. Finally, the CATE $\tau(\boldsymbol{x};t^*)$ can be estimated by minimizing the objective function (\ref{loss:squared_error_censor}) with $Y^*$, $w^M$ and $w^C$ plugged in. 

All the meta-learners discussed above have established theoretical guarantees. \cite{KennedyDR} developed the smoothness requirement of the true CATE function and derived the convergence rate of the DR-learner. \cite{Dlearning_TianLu2014} provided the theoretical results for the regularized estimator of D-learner with an $L_1$ Lasso penalty and justified the optimal efficiency augmentation for different types of outcomes; \cite{DEAlearning2017} showed the detailed theoretical justifications for extending the D-learner to the DEA-learner and compared its statistical properties with alternative methods; \cite{Rlearner2021} developed a quasi-oracle error bound for the R-learner. More details about the theoretical properties and proofs can be found in the respective original articles.

\subsection{Interpretable HTE estimation framework}
\label{sec: method_interpretable_ITE}

Suppose that we have $K$ candidate subgroups, $r_1(\boldsymbol{x}), r_2(\boldsymbol{x}),\cdots,r_K(\boldsymbol{x})$ that may potentially contribute to CATE prediction. We can re-parameterize $\tau(\boldsymbol{x};t^*)$ in the objective function (\ref{loss:squared_error_censor}) as $\tau(\boldsymbol{x};t^*)=\beta_0+\sum_{k=1}^K \beta_k r_k(\boldsymbol{x})$ for $k=1,2,\cdots,K$. Thus, our proposed interpretable framework estimates the CATE in survival outcomes by minimizing the following objective function with a penalty term to select important subgroups:
\begin{eqnarray}
\label{eq:objective_func_ipcw_complete_subgroup}
    \operatorname*{argmin}_{\boldsymbol{\beta}} \frac{1}{n^o} \sum_{i=1}^{n^o} w_i^C w_i^M\{Y_i^*-[\beta_0+\sum_{k=1}^K \beta_k r_k(\boldsymbol{x}_i)]\}^2 + \lambda \mathcal{R}(\boldsymbol{\beta}),
\end{eqnarray}
where $r_k(\boldsymbol{x}_i) \in \{0,1\}$ denotes whether a subject $i$ with covariate $\boldsymbol{x}_i$ falls into the $k$-th candidate subgroup or not; $\beta_k$ denotes the coefficient of the corresponding candidate subgroup $r_k(\boldsymbol{x}_i)$; $\mathcal{R}(\boldsymbol{\beta})$ is a penalty term for subgroup selection with a positive $\lambda$ as the tuning parameter.

\begin{algorithm}[ht]
\caption{An interpretable HTE estimation framework for survival outcomes}
\label{algorithm:framework_subgroup}

\textbf{Input:} Covariate $\boldsymbol{X}$, treatment $A$, pre-specified time $t^*$, observed outcome $I(T>t^*)$.

\textbf{Step 1: Construct $Y^*$, $w^M$ and $w^C$} \\
Estimate the following nuisance parameters on the training data.\\
(1) Estimate propensity score $e(\boldsymbol{X})$ and outcome regression models $S_a(t^*|\boldsymbol{X})$ or $S(t^*|\boldsymbol{X})$.\\
(2) Construct pITE $Y^*$ for individuals whose observed outcome $I(T>t^*)$ can be determined, using the selected meta-learner. \\
(3) Construct the meta-learner method-specific weight $w^M$.\\
(4) Estimate IPCW $w^C=\frac{1}{P(C>c|\boldsymbol{X}=\boldsymbol{x},A=a)}$ using all individuals and predict $w^C$ for each individual whose $I(T>t^*)$ can be determined.\\

\textbf{Step 2. Candidate subgroup generation}\\
(1) Minimize the objective function:
\begin{eqnarray}
\operatorname*{argmin}_\tau \frac{1}{n^o} \sum_{i=1}^{n^o} w_i^C w_i^M\left(Y_i^*-\tau(\boldsymbol{X}_i;t^*)\right)^2 \nonumber
\end{eqnarray}
through an ensemble tree-based algorithm \\
(2) Extract all terminal and internal nodes from the generated trees.\\
(3) Remove complementary and repeated tree nodes to obtain candidate subgroups $r_1(\boldsymbol{x}),r_2(\boldsymbol{x}), \cdots, r_K(\boldsymbol{x})$.\\

\textbf{Step 3. Subgroup selection} \\
(1) Transform the original covariates $\boldsymbol{X} \in \mathbb{R}^p$ into binary indicators $r_k(\boldsymbol{x}) \in \{0, 1\}$, with $r_k(\boldsymbol{x})$ from Step 2. \\
(2) Minimize the following objective function through a penalized regression:
\begin{eqnarray}
    \operatorname*{argmin}_{\boldsymbol{\beta}} \frac{1}{n^o} \sum_{i=1}^{n^o} w_i^C w_i^M\{Y_i^*-[\beta_0+\sum_{k=1}^K \beta_k r_k(\boldsymbol{x}_i)]\}^2 + \lambda \mathcal{R}(\boldsymbol{\beta}),  \nonumber
\end{eqnarray}
where $\lambda$ is a tuning parameter and $\mathcal{R}(\boldsymbol{\beta})$ is a penalty term.\\

\textbf{Output:} $L$ numbers of selected subgroups ($L<K$), denoted as $r_1(\boldsymbol{x}), r_2(\boldsymbol{x}), \cdots, r_L(\boldsymbol{x})$, with estimated coefficients $\hat{\beta}_l$, $l=1,2,\cdots,L$. 

For a new data point with covariate $\boldsymbol{x}$, CATE is predicted as: 
$    \hat{\tau}(\boldsymbol{x};t^*)=\hat{\beta}_0+\sum_{l=1}^L \hat{\beta}_{l} r_{l}(\boldsymbol{x}). $

\end{algorithm}

The algorithm \ref{algorithm:framework_subgroup} describes the proposed interpretable HTE estimation framework for survival outcomes. Flowchart \ref{fig:example_rule_tree} visually illustrates each step of the algorithm. In Step 1, we construct $Y^*$, $w^M$, and $w^C$, which involves estimating the following nuisance parameters: the propensity score $e(\boldsymbol{X})$, the conditional survival function in each treatment arm $S_a(t^*|\boldsymbol{X})$) or the conditional survival function ignoring the treatment arm $S(t^*|\boldsymbol{X})$). Parametric or semi-parametric methods can be applied (e.g., random forests (RF) to estimate $e(\boldsymbol{X})$, random survival forests (RSF) to estimate $S_a(t^*|\boldsymbol{X})$ or $S(t^*|\boldsymbol{X})$). \cite{metalearner_surv_JDS} studied the performance of different machine learning and deep learning methods, including RSF, Bayesian accelerated failure time model (BAFT) \citep{BAFT} and deep survival neural network (DNNSurv) \citep{DNNSurv}, to estimate conditional survival functions $S_a(t|\boldsymbol{x})$. More details about the performance using different machine learning and deep learning methods can be found in their paper. The estimated nuisance parameters are then used to construct $Y^*$ and $w^M$ within the meta-learner selected by the user. Finally, the censoring probability can be estimated using the Kaplan-Meier estimator (KM) if censoring is completely independent of survival time (i.e., $T \perp C$) or using RSF if censoring is covariate-dependent (i.e., $T \not\perp C$ but $T \perp C | \{\boldsymbol{X}, A$\}). Please note that if the sample size is large, one can split the data into a subset for estimating nuisance parameters and another subset for estimating the CATE function (with estimated nuisance parameters plugged in), as implemented in \cite{KennedyDR}. If the sample size is moderate or scarce, one can use cross-fitting or out-of-bag (OOB) prediction (without splitting the data) to estimate the nuisance parameters, as recommended by \cite{DoubledebiasedML_Chernozhukov2018}.


In Step 2, we plug in $Y^*$, $w^C$, and $w^M$ into the objective function (\ref{loss:squared_error_censor}) to solve for CATE using a tree-based method. This will transform the original covariate space $\boldsymbol{X} \in \mathbb{R}^p$ into candidate subgroups $\boldsymbol{r}(\boldsymbol{x})$. Various tree-based methods can be considered. For example, \cite{PRE_Rpackage} gives examples using CART and conditional inference tree (CTree), each offering distinct advantages. CART is a well-understood and widely implemented tree-based method in many \textit{R} packages; on the other hand, CTree addresses the variable selection bias by implementing unbiased splitting criteria that evaluate whether each covariate carries information about the outcome when determining the splitting point at internal nodes. Ensemble algorithms (e.g., gradient boosting, RF) can be used to extract tree nodes that collectively cover the covariate space as fully as possible. In the resulting trees, each internal or terminal node represents a potential candidate subgroup. Because complementary subgroups are perfectly correlated (i.e., $r_k(\boldsymbol{x})+r_{k^\prime}(\boldsymbol{x})=1$ for $k \neq k^\prime$), one subgroup from each complementary pair is removed. Duplicate subgroups generated across multiple trees are also eliminated. Figure \ref{fig:example_rule_tree}B shows an example of candidate subgroups generated from trees, represented as the colored nodes.

In Step 3, we transform the original covariates into binary indicators $r_k(\boldsymbol{x}) \in \{0, 1\}$, where each $r_k(\boldsymbol{x})$ corresponds to a candidate subgroup generated in Step 2. These indicators $r_1(\boldsymbol{x}),  r_2(\boldsymbol{x}), \cdots, r_K(\boldsymbol{x})$ are then used as new covariates in the objective function (\ref{eq:objective_func_ipcw_complete_subgroup}), where we solve for the coefficients $\beta$ by applying a regularization penalty term (e.g., Lasso, Elastic Net) with cross-validation (CV) to select the tuning parameter $\lambda$. For example, \cite{RuleFit2008} implemented the Lasso penalty, which yields a sparse set of selected subgroups and improves interpretability. In this paper, we also adopt the Lasso penalty to enhance the interpretability of HTE prediction. The resulting objective function in Step 3 becomes:
\begin{eqnarray}
\label{eq:objective_func_ipcw_complete_subgroup_lasso}
    \operatorname*{argmin}_{\boldsymbol{\beta}} \frac{1}{n^o} \sum_{i=1}^{n^o} w_i^C w_i^M\{Y_i^*-[\beta_0+\sum_{k=1}^K \beta_k r_k(\boldsymbol{x}_i)]\}^2 + \lambda\sum_{k=1}^K |\beta_k|.
\end{eqnarray}
The predicted CATE is $\hat{\tau}(\boldsymbol{x};t^*)=\hat{\beta}_0+\sum_{l=1}^{L}\hat{\beta}_lr_l(\boldsymbol{x})$, where $ r_l(\boldsymbol{x})$'s are final selected subgroups ($L < K$), determined based on the tuning parameter $\lambda$ value, which is choosen to make the best predictions through the CV. 

\section{Simulations}
\label{sec:simu}
In this section, we present comprehensive simulation studies conducted under various study designs, including both low-dimensional independent signal settings and high-dimensional correlated signal settings.

\subsection{Simulation designs}
\label{sec:simu_design}
We first considered an independent signal setting with ten independent covariates. $X_1, X_2, \cdots, X_5$ were binary covariates created by $X_j = I(\Tilde{X}_j > 0)$, where $\Tilde{X}_j$ was independently generated from a normal distribution $N(0,1)$, and $X_6, X_7, \cdots, X_{10}$ were continuous covariates independently generated from $N(0,1)$. 
Treatment assignment was determined by the propensity score with $\mathrm{logit}(e(\boldsymbol{X}))=0.4-0.3X_1-0.2X_6-0.3X_2-0.35X_7-0.2X_3-0.25X_8$. 

Potential survival times were simulated from a Weibull regression model: $T(a|\boldsymbol{X})=\lambda_a\{ \frac{-\mathrm{log}(U)}{\mathrm{exp}\{f_a(\boldsymbol{X})\}}\}^\frac{1}{\eta}$, where $U \sim \mathrm{Unif}[0,1]$ and $a=0$ or $1$. $f_a(\boldsymbol{X})=b(\boldsymbol{X})+a*h(\boldsymbol{X})$ introduced the source of HTE. $b(\boldsymbol{X})$ was a function of covariates shared between two treatment arms; $h(\boldsymbol{X})$ is another function of covariates only for the treatment arm. We considered simple to complex scenarios of $b(\boldsymbol{X})$ and $h(\boldsymbol{X})$ listed below. In Scenario 1, the covariates are in linear form and do not share between $b(\boldsymbol{X})$ and $h(\boldsymbol{X})$. In Scenario 2, we added one shared covariate in both $b(\boldsymbol{X})$ and $h(\boldsymbol{X})$. In Scenario 3, we considered the nonlinear form and/or interaction of the covariates in both $b(\boldsymbol{X})$ and $h(\boldsymbol{X})$. Under this setting, around half of the covariates are independent noise variables that do not contribute to either $b(\boldsymbol{X})$ or $h(\boldsymbol{X})$. Note that the subgroups are defined based on covariates in $b(\boldsymbol{X})$ and $h(\boldsymbol{X})$, where covariates in $h(\boldsymbol{X})$ directly contribute to CATE predictions, while those in $b(\boldsymbol{X})$ contribute indirectly. The shape parameter was set as $\eta=2$; scale parameter was set as $\lambda_0=16, 20, 20$ and $\lambda_1=26, 22, 22$ for scenarios 1 to 3 respectively. The censoring time was simulated independently from an exponential distribution to yield a 30\% censoring rate for each scenario. 

\begin{algorithm}
\caption*{\textbf{Simulation scenarios}}
\label{eq:scenarios}
     \textbf{S1:
     } 
     $b(\boldsymbol{X})=2X_6-1.2X_7$, \nonumber  

     \qquad $h(\boldsymbol{X})=2.8X_1-1.4X_2$. \nonumber

     \textbf{S2: 
     }
     $b(\boldsymbol{X})=1.6X_1-1.4X_6-1.2X_7$, \nonumber 

     \qquad $h(\boldsymbol{X})=2.5X_1-1.8X_2-2X_3$. \nonumber 

     \textbf{S3: 
     }
     $b(\boldsymbol{X})=1.6X_1-1.4X_6-1.2X_7-X_1X_7-0.8X_8^2$, \nonumber 

     \qquad $h(\boldsymbol{X})=2.5X_1-1.8X_2-2X_3-1.4X_1X_3$. \nonumber
\end{algorithm}

Additionally, we considered high-dimensional settings with correlated signals by generating covariates from a multivariate normal distribution $\Tilde{\boldsymbol{X}} \sim MVN(\boldsymbol{0},0.5\Sigma)$, where $\Sigma=\{\sigma_{jj^{\prime}} = e^{-|j-j^{\prime}|}, 1\leq j, j^{\prime} \leq p \}$ for $p=100$. 
$X_1, X_2, X_3, X_4, X_5, X_{11}, X_{12}, \cdots, X_{55}$ are binary covariates created by $X_j=I(\Tilde{X}_j > 0)$ for $j=1,2,3,4,5,11,12, \cdots,55$ and the rest are continuous with $X_j=\Tilde{X}_j$. We generated 100 training datasets with a sample size $n=1,000$ and one large test dataset with a sample size $N=10,000$.

In step 1, we applied RSF \citep{RSF} to estimate $S_a(t^*|\boldsymbol{X})$ and $S(t^*|\boldsymbol{X})$ and RF \citep{RF} to estimate $e(\boldsymbol{X})$. OOB predictions were used to predict the propensity score and conditional survival functions for each patient. The KM estimator with cross-fitting was used to estimate $P(C>\min(T,t^*))$ under this independent censoring setting. If censoring is potentially dependent on covariates, we recommend using model-based or machine learning approaches (e.g., RSF) to estimate the censoring probability. In Step 2, we implemented CTree with a gradient boosting algorithm to generate an ensemble of shallow trees through the R package \textit{pre}. CTree employs unbiased splitting criteria through a multiple testing procedure to select the variable to split at a node by testing whether the marginal distribution of the outcome equals to the conditional distribution of the outcome given a covariate \citep{conditional_inf_tree}. The maximum tree depth was set to 5, with a maximum of 500 trees generated. $\alpha$ in the multiple testing procedure (e.g., in Bonferroni correction) when splitting at a node was treated as a tuning parameter, as stated in the original paper \citep{conditional_inf_tree}. Based on our empirical experience, in the independent signal setting, we incorporated Bonferroni correction in CTree with $\alpha=0.1$, and no multiple testing adjustment was conducted for the high-dimensional correlated signal setting. In Step 3, we used the Lasso penalty and performed a 10-fold CV to select the tuning parameter $\lambda$ that yields the best predictions in terms of the smallest mean squared errors and output finally selected subgroups $r_1(\boldsymbol{x}), r_2(\boldsymbol{x}),\cdots,r_L(\boldsymbol{x})$.



Finally, we calculated subgroup importance score and variable importance score, denoted as: $R_l =|\hat{\beta}_l|\cdot\sqrt{s_l(1-s_l)}$ and $V_j = \sum_{X_j \in r_l} \frac{R_l}{c_l}$, where $s_l=\frac{1}{n^o}\sum_{i=1}^{n^o} r_l(\boldsymbol{x}_i)$ denotes the support of subgroup $r_l$ and $c_l$ denotes the number of subgroups that contain $X_j$. We also evaluated the prediction performance on test data based on the metrics introduced in Section \ref{sec:simu_eval}.

We compared the performance of DR-, DEA-, and R-learners with the scenarios when pITE $Y^*$ was generated directly using the predicted $\hat{\tau}(\boldsymbol{x};t^*)$ from CSF \citep{CSF_JRSSb2023} or BAFT. For CSF, we set the number of trees to be 500, mtry to be $p$ (the total number of covariates), and the minimum node size to be 15. For BAFT, we generated 500 trees with 100 MCMC iterations and obtained 1,000 posterior draws. In addition, we fit Cox models using the correct form of covariates with T-learner. Specifically, we fit separate Cox models for the treatment group and the control group. For each subject, we then predicted the survival probability under treatment and under control at time $t^*$. The CATE was computed as the difference between these two predicted survival probabilities. This Cox model result served as the ``true" model for benchmarking. 
In this paper, we estimated the CATE at the median survival time (i.e., $t^*$ = median survival time). Different time points may be selected depending on the specific research question.

In Section \ref{sec:simu_sensitivity_analysis}, we performed additional analysis under different censoring mechanisms (e.g., high censoring rates and covariate-dependent censoring scenarios). We also added a true null setting where there are no underlying subgroups.

\subsection{Evaluation metrics} 
\label{sec:simu_eval}
\textbf{Prediction performance.} 
We evaluated the prediction performance of each method in terms of Bias, binned-RMSE, and Spearman rank correlation, defined as:

\begin{eqnarray}
     \mbox{Bias} &=&\frac{1}{N}\sum\limits_{i=1}^{N}(\hat{\tau}_i(\boldsymbol{x}; t^*)-\tau_i(\boldsymbol{x}; t^*)), \quad \nonumber \\
     \mbox{binned-RMSE} &=& \frac{1}{Q}\sum\limits_{q=1}^{Q}\sqrt{\frac{1}{N_q}\sum\limits_{i=1}^{N_q}(\hat{\tau}_i(\boldsymbol{x}; t^*)-\tau_i(\boldsymbol{x}; t^*))^2},\quad \nonumber \\
     \mbox{Spearman Corr} &=& 1-\frac{6\sum d_i^2}{N(N^2-1)}, \nonumber 
\end{eqnarray}
where $\tau_i(\boldsymbol{x}; t^*)$ and $\hat{\tau}_i(\boldsymbol{x}; t^*)$ denote the true and predicted CATE value for subject $i$, respectively, and $d_i$ denotes the difference between the ranks of $\tau_i(\boldsymbol{x}; t^*)$ and $\hat{\tau}_i(\boldsymbol{x}; t^*)$ in the testing data. To evaluate treatment heterogeneity, we evaluated CATE predictions within each subgroup using the binned-RMSE \citep{HTE_mediansurv_ML_Hu2021}. We divided the predictions into $Q$ bins based on the order of true CATEs, calculated the MSE within each bin, and then averaged across all bins. We set $Q=50$ here.

In addition, we evaluated prediction accuracy in terms of treatment recommendation using the predicted CATE. The sign of true CATE values is assumed to be the gold standard: if $\tau(\boldsymbol{x};t^*)>0$, the subject should be recommended for treatment; if $\tau(\boldsymbol{x};t^*)\leq 0$, the subject should be recommended for control. If the subject's predicted CATE is greater than 0 (e.g., $\hat{\tau}(\boldsymbol{x};t^*)>0$), the subject was recommended to take treatment; if the subject's predicted CATE is less than or equal 0 (e.g., $\hat{\tau}(\boldsymbol{x};t^*)\leq0$), the subject was recommended to take control. Then, true positive (TP) is defined as the number of subjects who should take treatment and were also recommended for treatment; true negative (TN) is defined as the number of subjects who should take control and were also recommended for control; false positive (FP) is defined as the number of subjects who should take control but were recommended for treatment; false negative (FN) is defined as the number of subjects who should take treatment but were recommended to the control. Thus, we define accuracy as $\text{ACC}=\frac{\text{TP}+\text{TN}}{\text{N}}$ and F-score as $\frac{2\text{TP}}{2\text{TP}+FP+FN}$.

\vspace{1em}

\noindent \textbf{Subgroup identification accuracy.} We evaluated each method's ability to select the correct covariates that contribute to CATE prediction in either $b(\boldsymbol{X})$ or $h(\boldsymbol{X})$. Specifically, we recorded how often each covariate was used to define at least one selected subgroup across 100 simulation runs. In addition, we examined the total number of selected subgroups across these 100 simulations.

\subsection{Simulation results}
\subsubsection{Low-dimensional independent signal setting}
\label{sec:simu_results_lowdim}
\begin{figure}[h]
\centering
\includegraphics[width=4.5in]{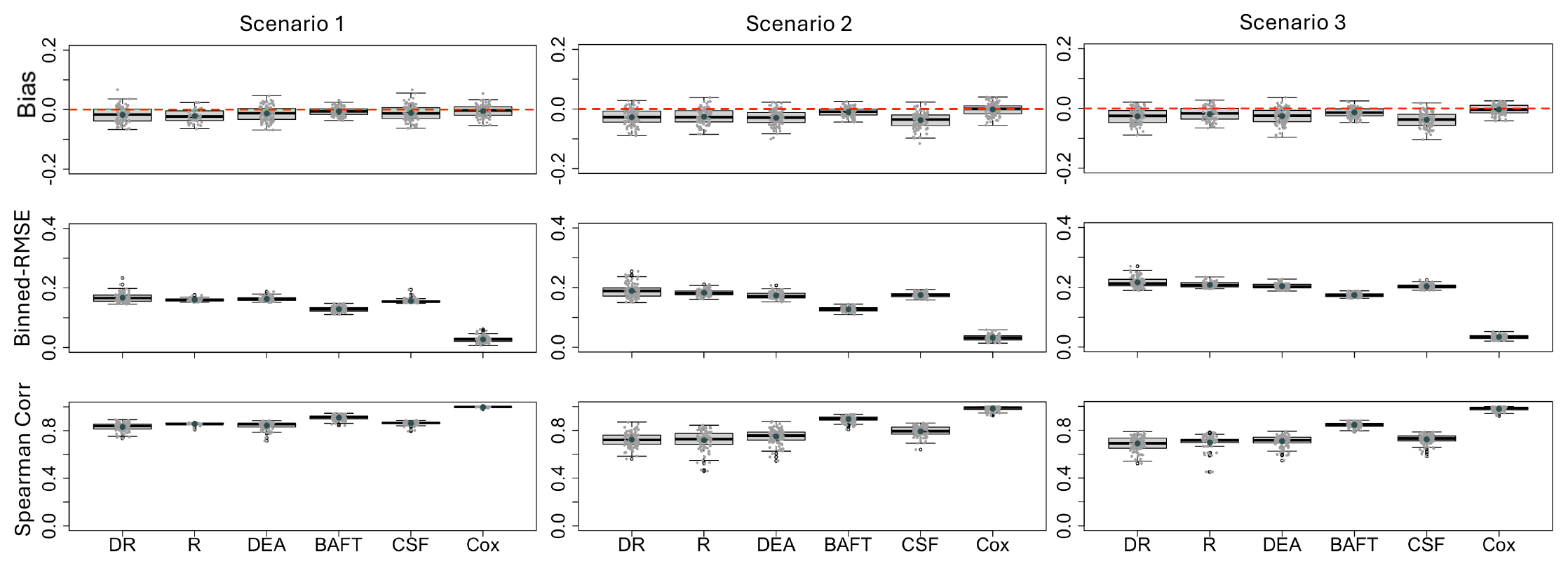}
\caption{Prediction accuracy under the independent signal setting with 30\% independent censoring. The prediction performance under the following meta-learners was examined: DR-learner (DR), R-learner(R), and DEA-learner (DEA). ``CSF'' represents causal survival forests. ``BAFT'' represents the Bayesian accelerated failure time model. ``Cox" represents the Cox model using correct form of covariates with T-learner. 
The upper panel shows biases, the middle panel shows binned-RMSEs, and the lower panel shows Spearman rank correlations.}
\label{fig:simu_box_lowdim_indepcen30_Bonf01}
\end{figure}

Figure \ref{fig:simu_box_lowdim_indepcen30_Bonf01} shows the prediction performance under the independent signal setting with Bonferroni corrections ($\alpha=0.1$) in CTree for generating candidate subgroups. In scenario 1, DEA-learner and BAFT produce minimal biases compared to other methods. In Scenarios 2 and 3, biases increase slightly for all methods, with the exception of BAFT, which continues to exhibit the smallest bias. Predictions based on CATEs from CSF produce the largest bias, particularly in complex scenarios (2 and 3). All methods show comparable RMSEs and high correlations with true CATEs, where BAFT slightly outperforms the others. Overall, when using three meta-learners to construct $Y^*$ under our framework, they show comparable prediction accuracy with the case when $Y^*$ is generated directly from the BAFT-predicted CATE.

The upper panel of Table \ref{tab: acc_F_IndepSig_HighdimSig_indepcen30_maintext} shows the overall accuracy (ACC) and the F-score in each scenario under the independent signal setting. In general, meta-learners achieve high accuracy and F-score in the independent signal setting and have comparable performance with BAFT and CSF. In scenario 1, DR-, R- and DEA-learners show higher ACC and F-score than BAFT with R-learner showing the highest ACC and F-score. CSF shows comparable ACC and F-score compared to the meta-learners. In Scenario 2, BAFT shows the highest ACC and F-score; CSF shows comparable ACC and F-score compared to the other three meta-learners. In Scenario 3, BAFT and the three meta-learners show comparable ACC and F-score, which are higher than CSF.

\begin{table}
\caption {Overall accuracy (ACC) and F-score with mean (SD) under the independent signal and high-dimensional signal settings with 30\% independent censoring.}
\centering
\label{tab: acc_F_IndepSig_HighdimSig_indepcen30_maintext}
\scalebox{0.88}{
\begin{tabular}{lcccccc}
\hline
                             & DR           & R             & DEA          & BAFT         & CSF          & Cox          \\ \hline
                             & \multicolumn{6}{c}{\textbf{Independent Signal Setting}}                                           \\ \hline
                             & \multicolumn{6}{c}{Scenario 1}                                                           \\ \hline
\multicolumn{1}{l|}{ACC}     & 92.34 (7.32) & 98.14 (6.49)  & 95.23 (9.97) & 90.48 (2.89) & 96.70 (8.30) & 99.40 (1.32) \\
\multicolumn{1}{l|}{F-score} & 92.85 (5.98) & 98.51 (5.15)  & 95.99 (7.30) & 90.88 (2.68) & 97.30 (6.08) & 99.42 (1.24) \\ \hline
                             & \multicolumn{6}{c}{Scenario 2}                                                           \\ \hline
\multicolumn{1}{l|}{ACC}     & 77.48 (4.57) & 79.91 (10.87) & 78.62 (6.57) & 85.85 (3.60) & 75.77 (7.77) & 96.08 (4.08) \\
\multicolumn{1}{l|}{F-score} & 82.51 (3.59) & 86.24 (6.53)  & 84.10 (4.63) & 88.86 (2.84) & 83.80 (4.32) & 96.77 (3.47) \\ \hline
                             & \multicolumn{6}{c}{Scenario 3}                                                           \\ \hline
\multicolumn{1}{l|}{ACC}     & 84.88 (5.04) & 85.53 (10.18) & 87.22 (7.28) & 87.59 (3.07) & 79.75 (4.91) & 96.44 (3.23) \\
\multicolumn{1}{l|}{F-score} & 90.00 (3.38) & 91.55 (5.85)  & 92.04 (4.44) & 91.61 (2.11) & 88.29 (2.62) & 97.57 (2.25) \\ \hline
\multicolumn{7}{c}{\textbf{High-dimensional Signal Setting}}                                                                     \\ \hline
                             & \multicolumn{6}{c}{Scenario 1}                                                           \\ \hline
\multicolumn{1}{l|}{ACC}     & 66.85 (2.70) & 97.64 (7.66)  & 85.29 (5.71) & 88.09 (3.02) & 53.75 (9.31) & 99.57 (1.17) \\
\multicolumn{1}{l|}{F-score} & 68.03 (2.79) & 98.14 (5.96)  & 86.23 (5.27) & 88.87 (2.66) & 67.46 (9.40) & 99.59 (1.10) \\ \hline
                             & \multicolumn{6}{c}{Scenario 2}                                                           \\ \hline
\multicolumn{1}{l|}{ACC}     & 62.86 (2.27) & 75.73 (11.31) & 72.14 (4.12) & 83.62 (2.49) & 70.18 (0.66) & 96.34 (4.37) \\
\multicolumn{1}{l|}{F-score} & 71.03 (2.17) & 83.52 (7.92)  & 79.19 (3.45) & 88.17 (1.89) & 82.38 (1.06) & 97.27 (3.39) \\ \hline
                             & \multicolumn{6}{c}{Scenario 3}                                                           \\ \hline
\multicolumn{1}{l|}{ACC}     & 66.81 (2.43) & 78.29 (7.75)  & 75.86 (4.89) & 84.82 (2.35) & 77.68 (0.00) & 96.23 (3.94) \\
\multicolumn{1}{l|}{F-score} & 76.42 (2.03) & 86.79 (5.22)  & 83.53 (3.55) & 89.93 (1.67) & 87.44 (0.00) & 97.46 (2.75) \\ \hline
\end{tabular}}
\end{table}

\begin{figure}
\centering
\includegraphics[width=4.5in]{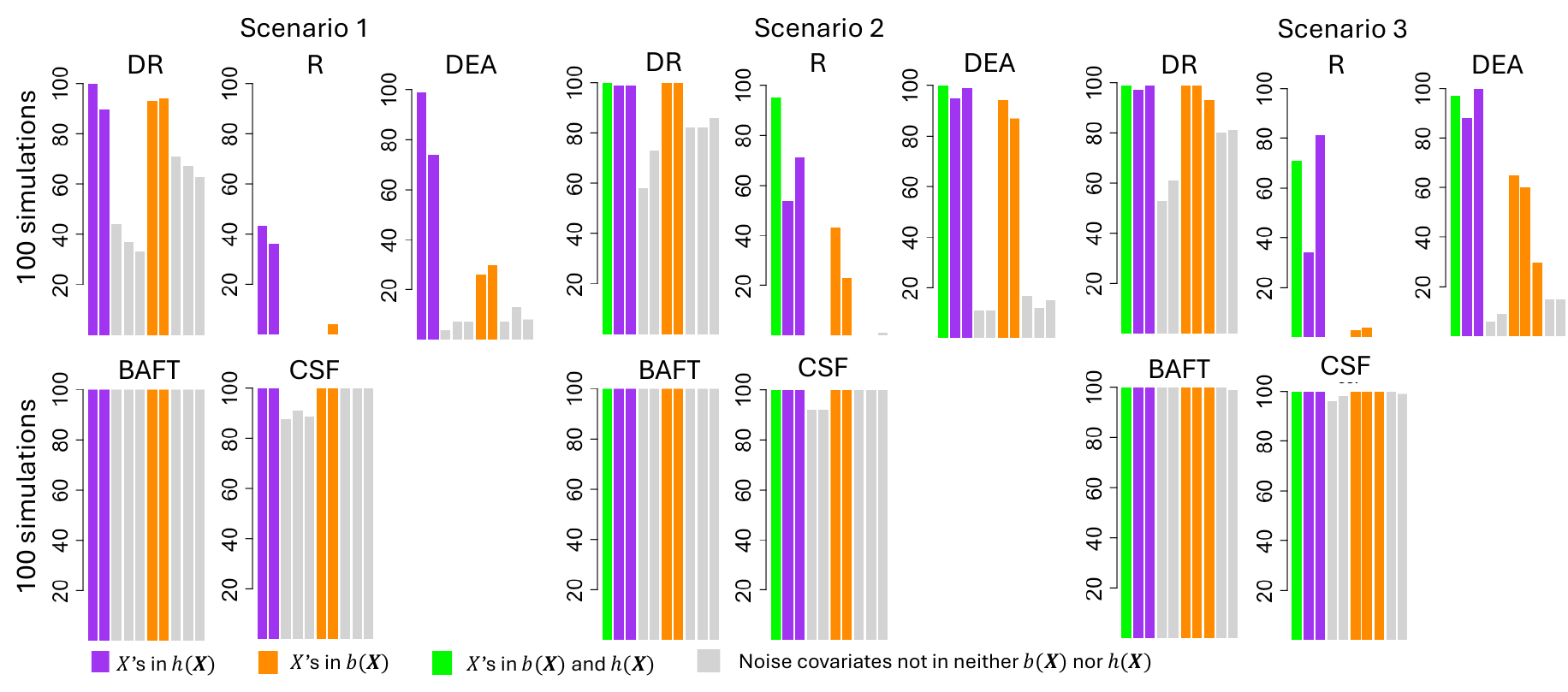}
\caption{Variable detection under the independent signal setting with 30\% independent censoring. The histogram shows the number of times each covariate is used to formulate any selected subgroups among 100 simulations. 
}
\label{fig:simu_hist_varfreq_lowdim_indepcen30_Bonf01}
\end{figure}

Figure \ref{fig:simu_hist_varfreq_lowdim_indepcen30_Bonf01} displays the frequency with which each covariate is included in at least one selected subgroup across 100 simulations. Under the DEA-learner, covariates in $h(\boldsymbol{X})$ (those that directly contributing to CATE) are selected mostly frequently to form subgroups (100\% for $X_1$, $>60\%$ for $X_2$), while covariates in $b(\boldsymbol{X})$ appear in about 40\% of simulations. Noise covariates (i.e., those not belonging to $h(\boldsymbol{X})$ or $b(\boldsymbol{X})$) are rarely selected with frequencies below 10\% across all three scenarios. The DR-learner tends to select noise covariates more frequently in all scenarios. Although BAFT achieves strong predictive performance, it includes all noise covariates in every simulation. The R-learner occasionally fails to generate more than one candidate subgroup, resulting in some simulations with no final subgroups identified. Thus, the frequencies of covariates used in generating subgroups under the R-learner are lower than in other methods. 

The upper panel of Table \ref{tab: simu_num_rules_selected_indep_signal10p_Bonf01} summarizes the number of selected subgroups under each method. The DEA-learner retains the sparseness, selecting 3 to 10 median number of subgroups among 100 simulations. The DR-learner also retains some sparseness but selects more subgroups than the DEA-learner (the median number of selected subgroups is between 11 and 24). BAFT and CSF select more than 100 subgroups that contain a lot of noise covariates. In addition, the effects of these selected subgroups under BAFT and CSF are minimal (i.e., $\hat{\beta}_k < 0.01$ for most coefficients). 

\begin{table}
\scriptsize
\caption{The number of selected subgroups under each method in the independent signal setting 
and the high-dimensional correlated signal setting 
among 100 simulations, median (min, max).}
\centering
\label{tab: simu_num_rules_selected_indep_signal10p_Bonf01}
\scalebox{1}{
\begin{tabular}{cccccc}
\hline
           & DR           & R        & DEA        & BAFT           & CSF            \\ \hline
           & \multicolumn{5}{c}{Independent Signals}                                \\ \hline
Scenario 1 & 11 (0, 35)    & 0 (0, 6) & 3 (0, 10)  & 153 (81, 203)  & 124 (72, 182)  \\
Scenario 2 & 24 (4, 65)   & 3 (0, 8) & 10 (3, 19) & 227 (166, 295) & 164 (106, 281) \\
Scenario 3 & 21 (0, 63)   & 2 (0, 6) & 7 (1, 14)  & 179 (136, 231) & 138 (96, 239) \\ \hline
           & \multicolumn{5}{c}{High-dimensional Correlated Signals}                  \\ \hline
Scenario 1 & 585 (403, 694) & 0 (0, 4) & 24 (3, 43)   & 460 (349, 530) & 383 (303, 616) \\
Scenario 2 & 623 (458, 735) & 2 (0, 7) & 26 (5, 53)  & 535 (418, 612) & 528 (368, 721) \\
Scenario 3 & 609 (435, 716) & 2 (0, 5) & 28 (3, 60)  & 480 (392, 588) & 449 (312, 648) \\ \hline
\end{tabular}}
\end{table}

\subsubsection{High-dimensional correlated signal setting}
\label{sec:simu_results_highdim}
\begin{figure}
\centering
\includegraphics[width=4.5in]{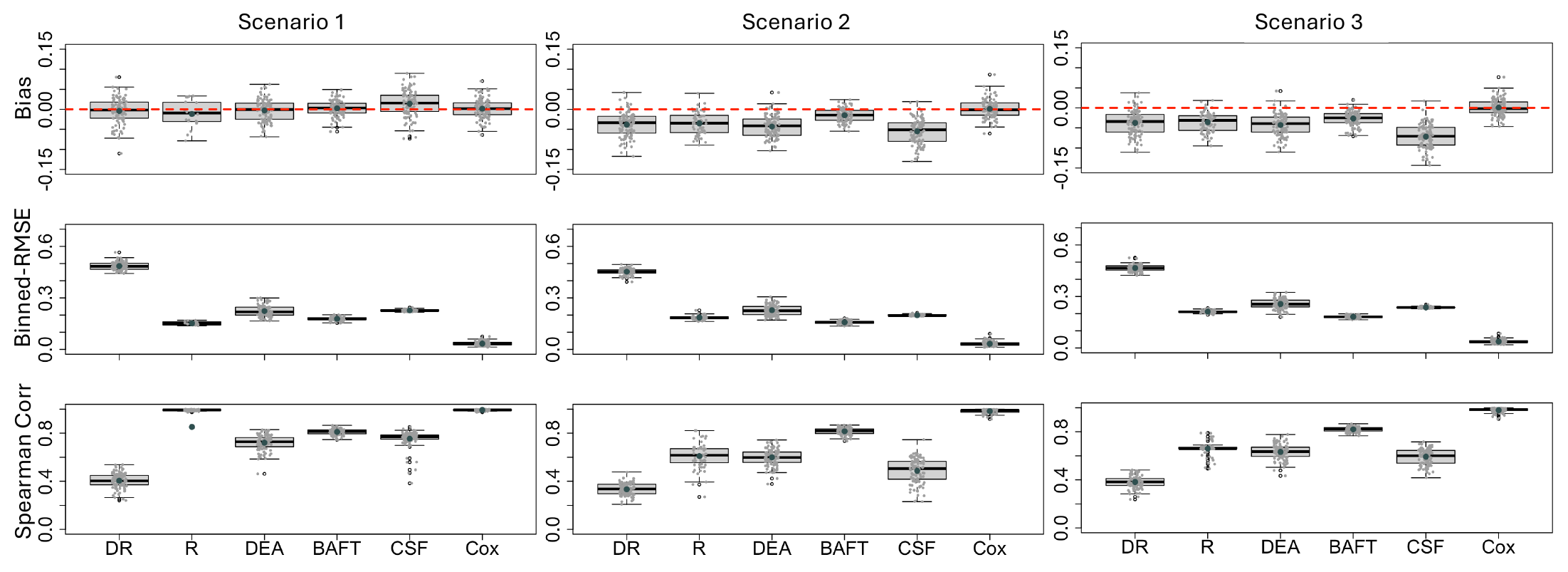}
\caption{Prediction accuracy under the high-dimensional correlated signal setting with a 30\% independent censoring rate. 
}
\label{fig:simu_box_highdim100p_indepcen30_Uni001}
\end{figure}

Figure \ref{fig:simu_box_highdim100p_indepcen30_Uni001} presents the prediction accuracy under the high-dimensional correlated signal setting with 30\% independent censoring. Unlike the low-dimensional independent signal setting, $\alpha$ was set as 0.01 without applying Bonferroni corrections when generating candidate subgroups using CTree. 
Consistent with the independent signal setting, both DR- and DEA-learners exhibit minimal bias in Scenario 1, with slightly increased bias in Scenarios 2 and 3. As in the low-dimensional setting, BAFT yields the smallest biases across all three scenarios. In contrast, CSF shows larger biases than the other methods, particularly in Scenarios 2 and 3. DR-learner also produces the highest RMSEs and the lowest correlations with the true CATE values across all scenarios, while the remaining methods demonstrate comparable RMSEs and correlations. 

The lower panel of Table \ref{tab: acc_F_IndepSig_HighdimSig_indepcen30_maintext} shows the overall accuracy (ACC) and F-score in each scenario in the independent signal setting. In general, all methods show lower accuracy and F-score compared to the independent signal setting with DR-learner showing the lowest values. R- and DEA-learners and CSF show comparable performance. BAFT shows the highest accuracy and F-score. However, its prediction performance is achieved by selecting hundreds of subgroups with very small effects, resulting in limited interpretability.

\begin{figure}[h]
\centering
\includegraphics[width=4.5in]{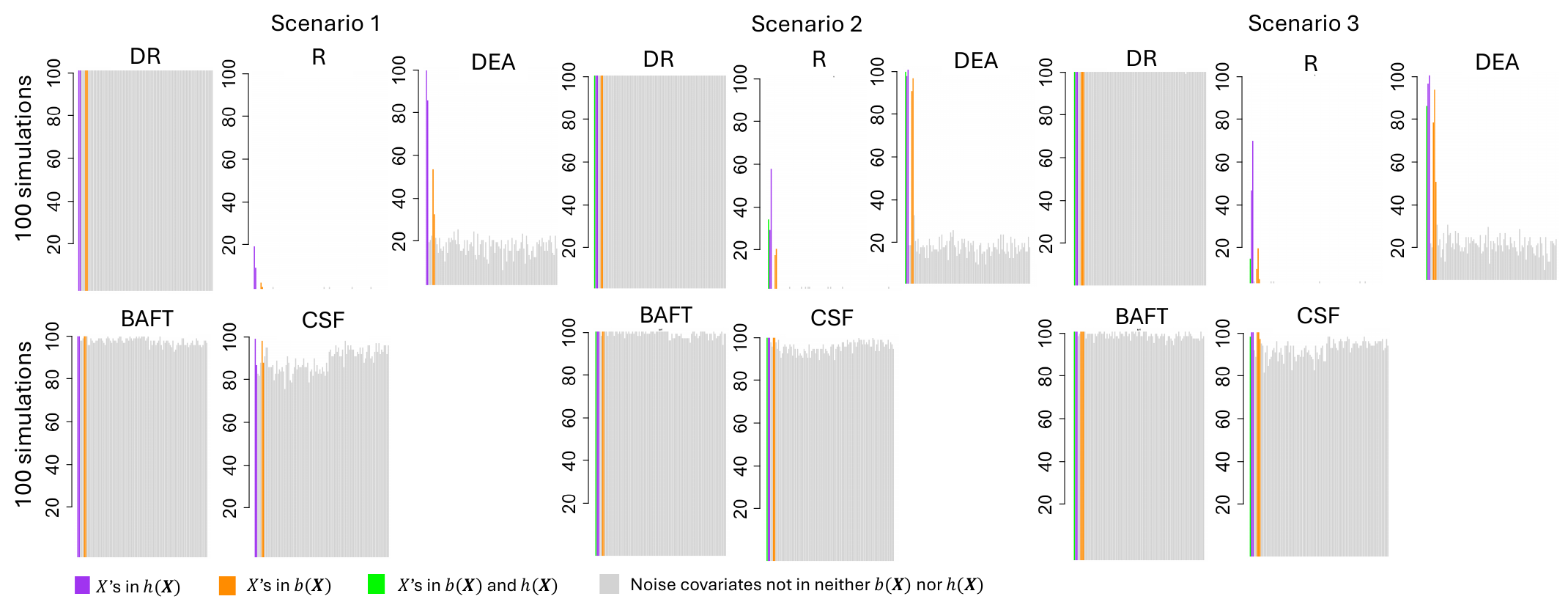}
\caption{Variable detection under the high-dimensional correlated signal setting with 30\% independent censoring. The histogram shows the number of times each covariate is ever used to formulate any selected subgroups among 100 simulations. 
}
\label{fig:simu_hist_varfreq_highdim100p_indepcen30_Uni001}
\end{figure}

Figure \ref{fig:simu_hist_varfreq_highdim100p_indepcen30_Uni001} shows the number of times each covariate is used to formulate any selected subgroup in 100 simulations. Similarly to the independent signal setting, the DEA-learner selects covariates in $h(\boldsymbol{X})$ to form subgroups in more than 80\% of simulations. In contrast, noise covariates are infrequently used, with low selection frequencies (around 20\%) across all scenarios. The DEA-learner also selects covariates in $b(\boldsymbol{X})$ with moderate frequencies. In comparison, DR-learner, BAFT, and CSF include a substantial number of noise covariates when forming subgroups. Notably, the R-learner only generates one candidate subgroup in 80\%, 40\%, and 30\% simulations from Scenarios 1 to 3, respectively. These runs were excluded from the evaluation since the Lasso penalty requires at least two covariates as input. Although the R-learner rarely selects noise variables, it also fails to consistently identify signal variables. As shown in the lower panel of Table \ref{tab: simu_num_rules_selected_indep_signal10p_Bonf01}, the DEA-learner maintains sparsity in the selected subgroups, while DR-learner, BAFT, and CSF each yield more than 380 subgroups with small effect sizes. The R-learner fails to generate subgroups in many simulation runs, with a median number of selected subgroups ranging from 0 to 2. 

\subsubsection{Additional simulation analysis}
\label{sec:simu_sensitivity_analysis}
\textbf{True null setting.} We conducted simulations under a true null setting where there is no underlying subgroup. We set $b(\boldsymbol{X})=0$ and $h(\boldsymbol{X})=0$ with true CATE values for all subjects being the same (i.e., $\tau(\boldsymbol{x};t^*)=0.21$). The censoring rate is still 30\%. Ten independent covariates from the independent signal setting in Section \ref{sec:simu_results_lowdim} were used as predictors for each method to be evaluated. The tuning parameters were set to be the same as those in Section \ref{sec:simu_results_lowdim}. In R-learner, there are about 5\% simulations with only one candidate subgroup generated and no candidate subgroup generated in the rest of the simulations. In DEA-learner, there are about 20\% simulations with only one candidate subgroup generated, 10\% simulations with no candidate subgroup generated, and no subgroup selected after penalization in the majority of the rest of 70\% simulations. DR-learner, BAFT and CSF generate $\geq2$ candidate subgroups in all 100 simulations; 65\% simulations for DR-learner have no subgroup selected after penalization; BAFT and CSF produce 340 and 374 selected subgroups on average after penalization. This validates our proposed algorithm, using pseudo-ITE from meta-learners to construct $Y^*$. Specifically, when there is no underlying subgroup, the method either fails to generate more than one candidate subgroup or yields no selected subgroup after penalization when using pseudo-ITE from the meta-learners. This is reflected in Web Figure 1 by plotting the frequency that each variable is ever used. DR-learner, BAFT and CSF detect a lot of noise variables. However, R- and DEA-learners rarely detect any variables. To evaluate the prediction performance due to large number of simulations with one candidate subgroup generated or no candidate subgroup generated, we fit a linear regression model if only one candidate subgroup was generated. Web Figure 2 shows the prediction accuracy (Spearman's rank correlations were not calculated since CATEs for all subjects are the same). DR-, R- and DEA-learners show minimal biases and comparable RMSEs. However, BAFT shows obvious bias with a larger RMSE compared to meta-learners. Although CSF shows minimal bias and the smallest RMSE, it also yields $>300$ subgroups selected, which violates the setup of the true null setting.

\textbf{Different censoring mechanism.} We conducted simulations considering different censoring situations for the low-dimensional independent signal setting. First, we varied the censoring rate. Web Figure 3 shows the prediction accuracy with 60\% censoring under the independent signal setting, while Web Figure 2 presents the variable detection results. The upper panel of Web Table 1 displays the number of selected subgroups. We observe similar prediction accuracy as seen in the low censoring rate setting. However, the DEA-learner, which typically shows low frequencies in selecting noise variables, demonstrates higher frequencies in selecting noise variables in the 60\% censoring case (Web Figure 4). The DEA-learner selected a median of 7 to 14 subgroups after Lasso penalization, which is slightly more than that in the low censoring rate setting. Other methods perform similarly to the low censoring setting. We also considered a covariate-dependent censoring setting. Specifically, censoring times were generated from a Weibull model: $C= \frac{-2log(U)}{\exp(b_0+b_1X_1+b_2X_2)}$, where $U \sim \mathrm{Unif}[0,1]$, and values of $b$'s are chosen to yield a 30\% censoring rate for Scenarios 1 to 3 respectively. We observe similar prediction accuracy (Web Figure 5), variable selection patterns (Web Figure 6), and numbers of selected subgroups (middle panel of Web Table 1) compared to the independent censoring setting with the same censoring rate.

\textbf{Different multiplicity adjustments.}
In high-dimensional correlated settings, we also conducted a simulation applying the Bonferroni correction with a larger $\alpha = 0.2$ in CTrees. This results in approximately 5\% of simulations with no subgroups selected, even though the DEA-learner still demonstrates the best overall performance (see Web Figures 7 and 8, and lower panel of Web Table 1 in the supplemental materials for details). Overall, using multiple testing adjustment generated fewer candidate subgroups which may not cover the covariate space well. For example, the R-learner failed to select any subgroups (lower panel of Web Table 1). Therefore, in high-dimensional correlated scenarios, we recommend avoiding multiple testing corrections and instead using a smaller significance threshold (e.g., $\alpha=0.01$).

\textbf{Data splitting.}
We also evaluated the performance of the proposed algorithm under data splitting. Training data were randomly divided into two folds: one for nuisance parameter estimation and one for candidate subgroup generation and selection. Web Figures 9 and 10 show the prediction performance and the frequency with which each covariate is included in at least one selected subgroup in 100 simulations under the independent signal setting. Web Table 2 shows the number of selected subgroups under each method. We observed similar performance compared to the results without data splitting and using cross-fitting (or OOB prediction) for nuisance parameter estimation. This aligns with the results in \cite{curth2021nonparametric} using cross-fitting for estimating CATE through meta-learners.

In summary, the DEA-learner consistently performs best in selecting sparse subgroups that include the correct covariates contributing to CATE, while also maintaining strong predictive accuracy across different simulation settings. In the following real data application, we use DEA-learner to perform the analysis.

\section{Real data application}
\label{sec:realdat}

\subsection{Data description}
\label{sec:realdat_data}
In this study, we analyzed data from two clinical trials, AREDS and AREDS2. The AREDS trial served as our training dataset, and the subsequent AREDS2 trial served as the test data. We analyzed 806 patients in AREDS data who were diagnosed with AMD but free of late AMD in at least one eye at enrollment, categorized as AMD stages 2, 
3, 
or 4. 
As shown in Table \ref{tab:AREDS_table_balanced}, patients were randomized based on their baseline disease severity and followed for up to 12 years, with 415 (51.49\%) randomized to the AREDS formula arm (treatment) and 391 (48.51\%) to the placebo arm (control).
The study population had a mean age of 68.77 years (SD = 5.05) at enrollment, 466 (57.82\%) female, and 393 (48.76\%) non-smokers. Patients included in AREDS had mild-to-moderate AMD, with approximately half (56.70\%) of the patients having moderate AMD and a mean baseline AMD severity scale of 4.09 (SD = 2.06). In addition to baseline characteristics, we included 46 SNPs (denoted as ``CE4'' SNPs), which were identified to be associated with treatment efficacy by the CE4 method \citep{CE4_survival} and 67 prognostic SNPs, which were reported to be associated with AMD progression \citep{Yan2017_AMD} (top SNPs with $p < 10^{-5}$).

\begin{table}
\caption {Baseline characteristics of the AREDS and AREDS2 data used in the analysis.}
\label{tab:AREDS_table_balanced}
\centering

\scalebox{0.63}{
\begin{tabular}{cccccc}
\hline
\multicolumn{1}{l}{}                                                                                                                                   & \multicolumn{4}{c}{AREDS trial}                                                                                                                                                                                                                                 & AREDS2 trial                                                              \\ \hline
Number of subjects                                                                                                                                     & \begin{tabular}[c]{@{}c@{}}All \\ ($n=806$)\end{tabular} & \begin{tabular}[c]{@{}c@{}}Placebo\\ ($n=391$)\end{tabular} & \begin{tabular}[c]{@{}c@{}} AREDS formula \\ ($n=415$)\end{tabular} & \textit{p}-value$^\star$ & \begin{tabular}[c]{@{}c@{}}AREDS formula \\ ($n=112$)\end{tabular} \\ \hline
\multicolumn{1}{l}{Age}                                                                                                                                &                                                          &                                                             &                                                                                            & 0.4905                                    &                                                                           \\
Mean (SD)                                                                                                                                              & 68.77 (5.05)                                             & 68.90 (5.17)                                                & 68.66 (4.93)                                                                               &                                           & 71.29 (7.25)                                                              \\
Median (Range)                                                                                                                                         & 68.60 (55.30-81.00)                                      & 68.50 (55.30-81.00)                                         & 68.70 (55.50-79.50)                                                                        &                                           & 71.00 (53.00, 86.00)                                                      \\
\multicolumn{1}{l}{Sex (n, \%)}                                                                                                                        &                                                          &                                                             &                                                                                            & 0.8236                                    &                                                                           \\
Female                                                                                                                                                 & 466 (57.82)                                              & 224 (57.29)                                                 & 242 (58.31)                                                                                &                                           & 50 (44.64)                                                                \\
Male                                                                                                                                                   & 340 (42.18)                                              & 167 (42.71)                                                 & 173 (41.69)                                                                                &                                           & 62 (55.36)                                                                \\
\multicolumn{1}{l}{Education (n, \%)}                                                                                                                        &                                                          &                                                             &                                                                                            & 0.3702                                    &                                                                           \\
$\leq$ High School                                                                                                                                                & 269 (33.37)                                              & 124 (31.71)                                                 & 145 (34.94)                                                                                &                                           & 27 (24.32)                                                                \\
College or above                                                                                                                                                   & 537 (66.63)                                              & 267 (68.28)                                                 & 270 (65.06)                                                                                &                                           & 84 (75.68)                                                                \\
\multicolumn{1}{l}{Smoking (n, \%)}                                                                                                                    &                                                          &                                                             &                                                                                            & 0.6877                                    &                                                                           \\
Never Smoked                                                                                                                                           & 393 (48.76)                                              & 194 (49.62)                                                 & 199 (47.95)                                                                                &                                           & 44 (39.29)                                                                \\
Former/Current Smoker                                                                                                                                  & 413 (51.24)                                              & 197 (50.38)                                                 & 216 (52.05)                                                                                &                                           & 68 (60.71)                                                                \\
\multicolumn{1}{l}{Baseline AREDS AMD severity score}                                                                                                  &                                                          &                                                             &                                                                                            & 0.6303                                    &                                                                           \\
Mean (SD)                                                                                                                                              & 4.09 (2.06)                                              & 4.13 (2.06)                                                 & 4.06 (2.07)                                                                                &                                           & 5.43 (0.96)                                                                          
\\ \hline
\multicolumn{5}{l}{{\small $^\star$ $p$-value is based on two-sample $t$-test or Chi-square test for continuous or categorical variables in AREDS trial.  
}}\\ & \multicolumn{1}{l}{}                                     & \multicolumn{1}{l}{}                                        & \multicolumn{1}{l}{}                                                                       & \multicolumn{1}{l}{}                      & \multicolumn{1}{l}{}  
\end{tabular}}

\end{table}

AREDS2 was a subsequent clinical trial that enrolled patients with more advanced AMD and followed them for six years. It evaluated the efficacy of four treatment arms: three modified versions of the original AREDS formula and one arm using the original formula itself. For this analysis, we included 112 patients from the AREDS formula arm whose baseline AMD severity was comparable to that of the AREDS participants. These patients had a mean baseline severity score of 5.43 (SD = 0.96), a mean enrollment age of 71.29 years (SD = 7.25), with 50 (44.64\%) female and 44 (39.29\%) non-smokers (see the right column of Table \ref{tab:AREDS_table_balanced}). After excluding two patients with missing SNP data, 110 patients were included in the final analysis.

Given both studies are RCTs, standard causal assumptions are typically satisfied by design (e.g., consistency and unconfoundedness). We further examined the propensity score distribution between the two treatment groups by fitting random forests using the SNPs and baseline covariates included in our analysis. The resulting propensity score distributions show good overlap between the treatment groups (Web Figure 11), supporting the population overlap assumption. In addition, assuming conditional independent censoring is reasonable here, as this assumption has been adopted in all prior survival analyses of the AREDS and AREDS2 datasets.

\subsection{Interpretable prediction of CATE for AREDS formula}
\label{sec:realdat_subgroup}
In this section, we applied the proposed interpretable HTE estimation framework to the AREDS clinical trial to estimate the CATE for delaying progression to late AMD. The time point of interest was five years, a clinically meaningful milestone also used in \cite{metalearner_surv_JDS}. We used the DEA-learner to construct pseudo-ITE, as it demonstrated the best performance under high-dimensional correlated signal settings in our simulation studies. RSF \citep{RSF} and RF \citep{RF} were performed to estimate $S_A(t^*|\boldsymbol{X}) \ (A=0, \ 1)$ and $e(\boldsymbol{X})$, respectively, using the same setup of tuning parameters as in the simulations. The model was trained on the AREDS dataset and evaluated on the AREDS2 dataset. To generate candidate subgroups, we implemented CTree with a gradient boosting algorithm, consistent with the approach used in the simulations. We set $\alpha=0.01$ without Bonferroni adjustment, following the same parameter settings and recommendations derived from our simulation results.

\begin{figure}
\centering
\includegraphics[width=4.5in]{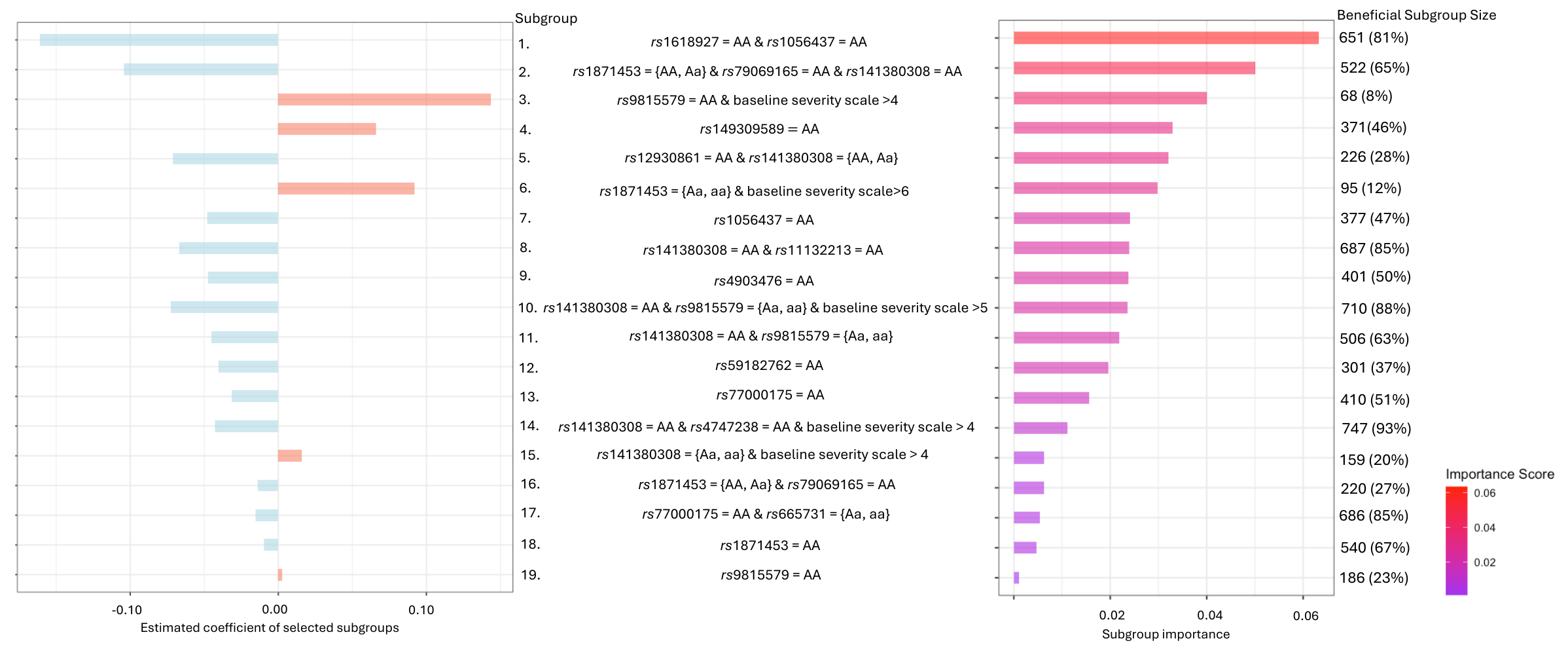}
\caption{Subgroups identified by training the algorithm on AREDS at year five. The left panel shows the estimated coefficients of selected subgroups. The right panel shows the subgroup importance ordered by the subgroup importance score. Subgroup size of the beneficial subgroup in AREDS is listed in the right column. 
}
\label{fig:AREDS_AREDS2_rule_imp_yr5_Uni001}
\end{figure}

Compared to previous studies on AMD \citep{CE4_survival,Yan2017_AMD,pseudoreg_survival_Bo2024}, our approach identified subgroups that interpret CATE predictions by capturing interaction effects between SNPs and/or baseline covariates. In contrast, prior work primarily focused on identifying individual SNPs through genome-wide association studies or post-hoc variable selection following CATE estimation using ``black-box'' machine learning methods. Figure \ref{fig:AREDS_AREDS2_rule_imp_yr5_Uni001} presents the subgroups identified by training the DEA-learner on the AREDS dataset using our proposed interpretable HTE estimation framework for survival outcomes. A total of nineteen subgroups were found to be predictive of CATE in delaying time-to-AMD progression. Fourteen SNPs and the baseline AMD severity scale formulate these nineteen subgroups, in which nine SNPs are CE4 SNPs (i.e., associated with treatment efficacy) with one from chromosome (CHR) 3 ($rs$9815579), one from CHR 5 ($rs$149309589), three from CHR 10 ($rs$1618927,$rs$1871453, $rs$4747238), three from CHR 14 ($rs$1056437, $rs$4903476, $rs$77000175) and one from CHR 19 ($rs$141380308); and five SNPs are prognostic SNPs (i.e., associated with disease progression regardless of the treatment), including $rs$59182762 (CHR 1), $rs$11132213 (CHR 4), $rs$79069165 (CHR 5), $rs$665731 (CHR 11), $rs$12930861 (CHR 16). Among the 19 identified subgroups, twelve are defined by interactions between two or more covariates. Subgroups with positive estimated coefficients (highlighted in orange) represent beneficial subgroups, indicating enhanced treatment effects. In contrast, subgroups with negative estimated coefficients (highlighted in blue) are considered adverse subgroups, where treatment effects are diminished or harmful. The complement of these adverse subgroups can also be interpreted as beneficial. The right column of Figure \ref{fig:AREDS_AREDS2_rule_imp_yr5_Uni001} displays the size of each subgroup within the AREDS trial population. It is important to note that the selected subgroups are not mutually exclusive. The primary goal of our proposed method is to enhance the interpretability of CATE predictions by uncovering interaction effects among covariates, rather than to identify all disjoint subgroups. External validation is necessary to assess the reproducibility of the identified subgroups. In the following section, we validate these results using the AREDS2 dataset.

\begin{figure}
\centering
\includegraphics[width=4.5in]{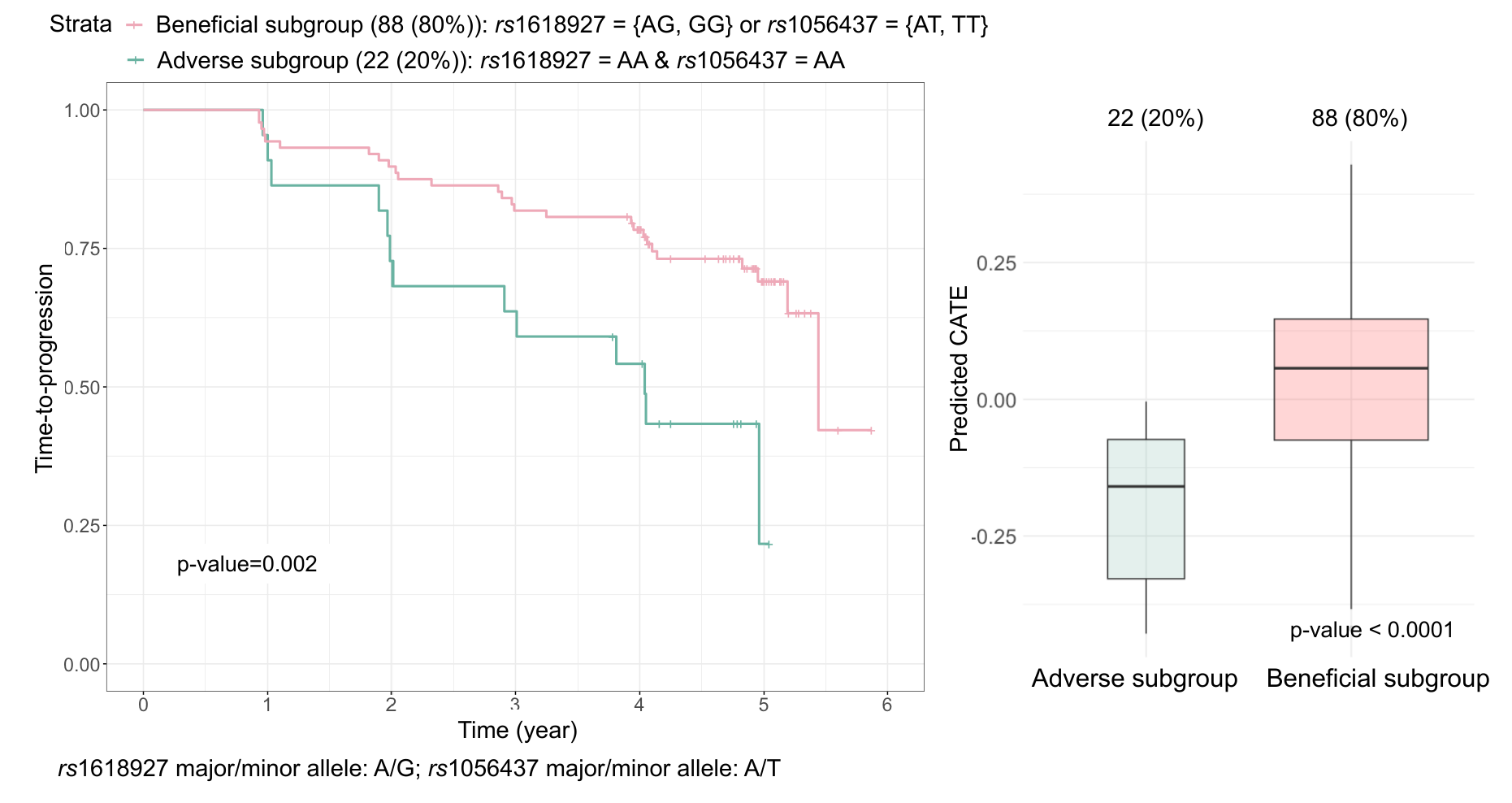}
\caption{Validation results of the top one selected subgroup on AREDS2. 
The left panel shows KM plots on AREDS2 separated by the two groups ($p$-value is from the log-rank test). 
The right panel shows the distributions of predicted CATEs on AREDS2 plotted by the two subgroups ($p$-value is from the Wilcoxon rank sum test). 
}
\label{fig:KM_box_AREDS2_rule1_yr5_Uni001_114SNP}
\end{figure}

In Figure \ref{fig:AREDS_AREDS2_rule_imp_yr5_Uni001}, the top-ranked subgroup exhibits a substantially higher subgroup importance score than all other selected subgroups. In the AREDS trial, this top beneficial subgroup, defined as ``$rs1618927=\{\text{AG}, \text{GG}\} \text{ or } rs1056437 = \{\text{AT}, \text{TT}\}$'', is the complement of its adverse counterpart,  ``$rs1618927 = \text{AA} \text{ and } \\ rs1056437 = \text{AA}$'', where the major/minor alleles for $rs1618927$ 
are A/G and for $rs1056437$ are A/T. This beneficial subgroup includes 651 patients (81\%) in the AREDS trial. We validated this subgroup in the AREDS2 trial, in which it comprised 88 patients (80\%). Among these patients, 58 (66\%) had positive CATE predictions. We plotted KM curves for the top beneficial subgroup versus its complementary adverse subgroup in AREDS2 (left panel of Figure \ref{fig:KM_box_AREDS2_rule1_yr5_Uni001_114SNP}) and conducted a log-rank test to assess the difference between the two groups. The KM curves are clearly separated, with a statistically significant difference (
log-rank test $p$-value = 0.002). Validation of the other selected subgroups can be performed similarly in the AREDS2 dataset.

We also compared the results from the DEA-learner to the results of BAFT. Web Figure 11 shows the top 20 selected subgroups by BAFT. BAFT selected a total of 367 subgroups with small effects, evidenced by the coefficients for the selected subgroups being much smaller compared to the DEA-learner (left panel of Web Figure 12). The top 1 and 2 beneficial subgroups define the subgroup using a high baseline AMD severity scale, resulting in a much smaller group size compared to the DEA-learner. The results cannot be validated on AREDS2, as there are no patients in AREDS2 falling into these two subgroups. We plotted KM curves and predicted CATE in ARED2 by the third top subgroup (Web Figure 13). The KM curves are not separable between the beneficial subgroup and its adverse subgroup, and the box plots of predicted CATE are more overlapping compared with the top one subgroup identified by the DEA-learner. 

\section{Conclusions and Discussions}
\label{sec:conclusion}
In this paper, we propose an interpretable framework for estimating CATE in survival outcomes with integrated causal subgroup discovery. The framework incorporates causal inference methods for CATE estimation, inverse probability of censoring weighting to handle right-censored data, and interpretable machine learning algorithms for subgroup identification. It offers flexibility in estimating nuisance parameters, choosing different meta-learners for constructing pseudo-ITEs, and applying tree-based methods for subgroup generation and selection via penalty terms. We evaluated the performance of three meta-learners, each combined with the conditional inference trees to simultaneously estimate CATE and identify subgroups. Among them, the DEA-learner demonstrated the highest subgroup identification accuracy while maintaining strong predictive performance. We applied the DEA-learner-based framework to the AREDS trial and identified genetic subgroups exhibiting treatment heterogeneity, with findings validated using the independent AREDS2 dataset.

One limitation of our proposed method is that the choice of the time of interest can influence the magnitude of estimated subgroup effects. Determining a meaningful time point requires prior knowledge or preliminary analysis. Furthermore, in implementing the conditional inference tree for unbiased splitting, it is necessary to decide whether to apply multiple testing adjustments and select a significance level ($\alpha$) for the splitting criterion. Although we provide recommendations based on simulation studies in various scenarios, the choice of $\alpha$ remains subjective. A stringent $\alpha$ level may lead to sparse subgroup identification and potentially overlook meaningful subgroups, while a more lenient $\alpha$ level may yield a larger number of subgroups, some with small effect sizes. Based on our real data analyses, we observed that the top subgroups, i.e., those with the highest variable importance scores, are generally consistent regardless of the chosen $\alpha$ level. We also recommend conducting additional survival analyses to further evaluate the effects of identified subgroups. 

Our framework can be extended in several important directions. The current method assumes no unmeasured confounding, which may limit its applicability in observational studies (e.g., electronic health records) unless a comprehensive set of covariates is available. A key direction for future work is therefore to adapt the proposed method to accommodate observational study designs. Second, interpretable neural network methods have emerged as a promising direction. For example, \cite{survRelu_2024_Sun} proposed the survReLU algorithm, which integrates a tree-based algorithm into neural networks for survival prediction. Incorporating such approaches into future work could be valuable, as neural networks can effectively capture complex data structures such as medical images. Finally, this paper has focused on the methodological development of interpretable CATE prediction. Although the three meta-learners used in generating pseudo-ITEs have established theoretical guarantees (see Section \ref{sec: method_pITE}), we did not provide theoretical results for the overall framework. Future work could investigate theoretical properties such as the convergence of CATE estimators and type I error control in subgroup identification. 

\subsection*{\small \textbf{Acknowledgements}}
The author(s) disclosed receipt of the following financial support for the research, authorship and/or publication of this article: This work was funded by [details omitted for blinded peer review].

\subsection*{\small \textbf{Supporting information}}
Supplemental material for this article is available online. The key functions for running simulations and real data analysis are available on GitHub and will be made available once the manuscript is accepted. 

%
\subsection*{\small \textbf{Conflict of interest}}
The authors declare that they have no conflict of interest.

\subsection*{\small \textbf{Data Availability Statement}}
Both phenotype and genotype data of AREDS and AREDS2 are available from the online repository dbGap (accession: phs000001.v3.p1 and phs001039.v1.p1, respectively).

\bibliographystyle{plainnat}
\bibliography{reference}

@article{rubin1974,
author = "Rubin, Donald B",
journal = "Journal of Educational Psychology",
number = "4",
pages = "688--701",
title = "Estimating causal effects of treatments in randomized and nonrandomized studies",
volume = "66",
year = "1974"
}

@article{Neyman1990,
author = "Splawa-Neyman, J. and Dabrowska, D.M. and Speed, T.",
journal = "Statistical Science",
number = "4",
pages = "465--472",
title = "On the Application of Probability Theory to Agricultural Experiments. Essay on Principles.",
volume = "5",
year = "1990"
}

@article{pseudoreg_survival_Bo2024,
author = {Bo, Na and Jeong, Jong-Hyeon and Forno, Erick and Ding, Ying},
title = {Evaluating Meta-Learners to Analyze Treatment Heterogeneity in Survival Data: Application to Electronic Health Records of Pediatric Asthma Care in COVID-19 Pandemic},
journal = {Statistics in Medicine},
volume = {44},
number = {3-4},
pages = {e10333},
doi = {https://doi.org/10.1002/sim.10333},
url = {https://onlinelibrary.wiley.com/doi/abs/10.1002/sim.10333},
eprint = {https://onlinelibrary.wiley.com/doi/pdf/10.1002/sim.10333},
year = {2025}
}

@article{HTE_mediansurv_ML_Hu2021,
author = {Hu, Liangyuan and Ji, Jiayi and Li, Fan},
title = {Estimating heterogeneous survival treatment effect in observational data using machine learning},
journal = {Stat Med},
volume = {40},
number = {21},
pages = {4691-4713},
doi = {https://doi.org/10.1002/sim.9090},
url = {https://onlinelibrary.wiley.com/doi/abs/10.1002/sim.9090},
eprint = {https://onlinelibrary.wiley.com/doi/pdf/10.1002/sim.9090},
year = {2021}
}

@article{KennedyDR,
author = {Edward H. Kennedy},
title = {Towards optimal doubly robust estimation of heterogeneous causal effects},
volume = {17},
journal = {Electron J Stat},
number = {2},
publisher = {Institute of Mathematical Statistics and Bernoulli Society},
pages = {3008 -- 3049},
year = {2023},
doi = {10.1214/23-EJS2157},
URL = {https://doi.org/10.1214/23-EJS2157}
}

@article{Dlearning_TianLu2014,
author = {Lu Tian and Ash A. Alizadeh and Andrew J. Gentles and Robert Tibshirani},
title = {A Simple Method for Estimating Interactions Between a Treatment and a Large Number of Covariates},
journal = {Journal of the American Statistical Association},
volume = {109},
number = {508},
pages = {1517-1532},
year  = {2014},
publisher = {Taylor & Francis},
doi = {10.1080/01621459.2014.951443},
    note ={PMID: 25729117},

URL = { 
    
        https://doi.org/10.1080/01621459.2014.951443
    
    

},
eprint = { 
    
        https://doi.org/10.1080/01621459.2014.951443
    
    

}

}

@article{DEAlearning2017,
author = {Chen, Shuai and Tian, Lu and Cai, Tianxi and Yu, Menggang},
title = {A general statistical framework for subgroup identification and comparative treatment scoring},
journal = {Biometrics},
volume = {73},
number = {4},
pages = {1199-1209},
doi = {https://doi.org/10.1111/biom.12676},
url = {https://onlinelibrary.wiley.com/doi/abs/10.1111/biom.12676},
eprint = {https://onlinelibrary.wiley.com/doi/pdf/10.1111/biom.12676},
year = {2017}
}

@article{Rlearner2021,
    author = {Nie, X and Wager, S},
    title = "{Quasi-oracle estimation of heterogeneous treatment effects}",
    journal = {Biometrika},
    volume = {108},
    number = {2},
    pages = {299-319},
    year = {2020},
    month = {09},
    issn = {0006-3444},
    doi = {10.1093/biomet/asaa076},
    url = {https://doi.org/10.1093/biomet/asaa076},
    eprint = {https://academic.oup.com/biomet/article-pdf/108/2/299/37938940/asaa076\_supplementary\_data.pdf},
}

@Article{RSF,
    title = {Random survival forests for R},
    author = {H. Ishwaran and U.B. Kogalur},
    journal = {R News},
    year = {2007},
    volume = {7},
    number = {2},
    pages = {25--31},
    month = {October},
    url = {https://CRAN.R-project.org/doc/Rnews/},
}

@article{RuleFit2008,
 ISSN = {19326157},
 URL = {http://www.jstor.org/stable/30245114},
 author = {Jerome H. Friedman and Bogdan E. Popescu},
 journal = {The Annals of Applied Statistics},
 number = {3},
 pages = {916--954},
 publisher = {Institute of Mathematical Statistics},
 title = {Predictive Learning via Rule Ensembles},
 urldate = {2023-03-12},
 volume = {2},
 year = {2008}
}

@article{RF,
    title = {Classification and Regression by randomForest},
    author = {Andy Liaw and Matthew Wiener},
    journal = {R News},
    year = {2002},
    volume = {2},
    number = {3},
    pages = {18-22},
    url = {https://CRAN.R-project.org/doc/Rnews/},
  }

@article{conditional_inf_tree,
author = {Torsten Hothorn and Kurt Hornik and Achim Zeileis},
title = {Unbiased Recursive Partitioning: A Conditional Inference Framework},
journal = {Journal of Computational and Graphical Statistics},
volume = {15},
number = {3},
pages = {651-674},
year  = {2006},
publisher = {Taylor & Francis},
doi = {10.1198/106186006X133933},

URL = { 
    
        https://doi.org/10.1198/106186006X133933
    
    

},
eprint = { 
    
        https://doi.org/10.1198/106186006X133933
    
    

}

}

@article{PRE_Rpackage,
 title={Fitting Prediction Rule Ensembles with R Package pre},
 volume={92},
 url={https://www.jstatsoft.org/index.php/jss/article/view/v092i12},
 doi={10.18637/jss.v092.i12},
 number={12},
 journal={Journal of Statistical Software},
 author={Fokkema, Marjolein},
 year={2020},
 pages={1–30}
}

@article{metalearner_surv_JDS,
    author = {Na Bo and Yue Wei and Lang Zeng and Chaeryon Kang and Ying Ding},
    title = {A Meta-Learner Framework to Estimate Individualized Treatment Effects for Survival Outcomes},
    journal = {Journal of Data Science},
    year = {2024},
    pages = {1--19},
    doi = {10.6339/24-JDS1119},
    issn = {1680-743X},
    publisher = {School of Statistics, Renmin University of China}
}

@article{RuleEnsemble_KeWan_SIM2023,
author = {Wan, Ke and Tanioka, Kensuke and Shimokawa, Toshio},
title = {Rule ensemble method with adaptive group lasso for heterogeneous treatment effect estimation},
journal = {Statistics in Medicine},
volume = {42},
number = {19},
pages = {3413-3442},
doi = {https://doi.org/10.1002/sim.9812},
url = {https://onlinelibrary.wiley.com/doi/abs/10.1002/sim.9812},
eprint = {https://onlinelibrary.wiley.com/doi/pdf/10.1002/sim.9812},
year = {2023}
}

@article{DNNSurv,
author = "Sun, Tao and Wei, Yue and Chen, Wei and Ding, Ying",
fjournal = "Statistics in Medicine",
journal="Statistics in Medicine",
volume = "39",
issue = "30",
pages = "4605--4620",
title = "Genome-wide association study-based deep learning for survival prediction",
year = "2020"}

@article{BAFT,
    author = {Henderson, Nicholas C and Louis, Thomas A and Rosner, Gary L and Varadhan, Ravi},
    title = "{Individualized treatment effects with censored data via fully nonparametric Bayesian accelerated failure time models}",
    journal = {Biostatistics},
    volume = {21},
    number = {1},
    pages = {50-68},
    year = {2018},
    month = {07},
    issn = {1465-4644},
    doi = {10.1093/biostatistics/kxy028},
    url = {https://doi.org/10.1093/biostatistics/kxy028},
    eprint = {https://academic.oup.com/biostatistics/article-pdf/21/1/50/40836273/biosts\_21\_1\_50\_s4.pdf},
}

@article{CSF_JRSSb2023,
    author = {Cui, Yifan and Kosorok, Michael R and Sverdrup, Erik and Wager, Stefan and Zhu, Ruoqing},
    title = "{Estimating heterogeneous treatment effects with right-censored data via causal survival forests}",
    journal = {Journal of the Royal Statistical Society Series B: Statistical Methodology},
    year = {2023},
    month = {02},
    issn = {1369-7412},
    doi = {10.1093/jrsssb/qkac001},
    url = {https://doi.org/10.1093/jrsssb/qkac001},
    note = {qkac001},
    eprint = {https://academic.oup.com/jrsssb/advance-article-pdf/doi/10.1093/jrsssb/qkac001/49531686/qkac001\_supplementary\_data.pdf},
}

@misc{curth2021nonparametric,
      title={Nonparametric Estimation of Heterogeneous Treatment Effects: From Theory to Learning Algorithms}, 
      author={Alicia Curth and Mihaela van der Schaar},
      year={2021},
      eprint={2101.10943},
      archivePrefix={arXiv},
      primaryClass={stat.ML}
}

@article{Yan2017_AMD,
author = {Yan, Qi and Ding, Ying and Liu, Yi and Sun, Tao and Fritsche, Lars G and Clemons, Traci and Ratnapriya, Rinki and Klein, Michael L and Cook, Richard J and Liu, Yu and Fan, Ruzong and Wei, Lai and Abecasis, Gonçalo R and Swaroop, Anand and Chew, Emily Y and {AREDS2} Research Group and Weeks, Daniel E and Chen, Wei},
title = "Genome-wide analysis of disease progression in age-related macular degeneration",
journal = "Human Molecular Genetics",
volume = "27",
number = "5",
pages = "929-940",
year = "2018"}

@article{CE4_survival,
author = {Yue Wei and Jason C. Hsu and Wei Chen and Emily Y. Chew and Ying Ding.},
journal = {Statistics in Medicine},

title = {Identification and Inference for Subgroups with Differential Treatment Efficacy from Randomized Controlled Trials with Survival Outcomes through Multiple Testing.},
volume = {40},
number = {29},
pages = {6523-6540},
year = {2021},
}

@article{DoubledebiasedML_Chernozhukov2018,
    author = {Chernozhukov, Victor and Chetverikov, Denis and Demirer, Mert and Duflo, Esther and Hansen, Christian and Newey, Whitney and Robins, James},
    title = "{Double/debiased machine learning for treatment and structural parameters}",
    journal = {The Econometrics Journal},
    volume = {21},
    number = {1},
    pages = {C1-C68},
    year = {2018},
    month = {01},
    issn = {1368-4221},
    doi = {10.1111/ectj.12097},
    url = {https://doi.org/10.1111/ectj.12097},
    eprint = {https://academic.oup.com/ectj/article-pdf/21/1/C1/27684918/ectj00c1.pdf},
}

@misc{CRE_havard2020,
      title={Causal Rule Ensemble: Interpretable Discovery and Inference of Heterogeneous Treatment Effects}, 
      author={Falco J. Bargagli-Stoffi and Riccardo Cadei and Kwonsang Lee and Francesca Dominici},
      year={2024},
      eprint={2009.09036},
      archivePrefix={arXiv},
      primaryClass={stat.ME},
      url={https://arxiv.org/abs/2009.09036}, 
      howpublished={}
}

@article{RuleEnsemblePrognostic_SMMR2024,
author = {Mayu Hiraishi and Ke Wan and Kensuke Tanioka and Hiroshi Yadohisa and Toshio Shimokawa},
title ={Causal rule ensemble method for estimating heterogeneous treatment effect with consideration of prognostic effects},
journal = {Statistical Methods in Medical Research},
volume = {33},
number = {6},
pages = {1021-1042},
year = {2024},
doi = {10.1177/09622802241247728},
    note ={PMID: 38676367},
URL = { https://doi.org/10.1177/09622802241247728},
eprint = {https://doi.org/10.1177/09622802241247728}
}

@misc{wan2023survivalcausalruleensemble,
      title={Survival causal rule ensemble method considering the main effect for estimating heterogeneous treatment effects}, 
      author={Ke Wan and Kensuke Tanioka and Toshio Shimokawa},
      year={2023},
      eprint={2309.11914},
      archivePrefix={arXiv},
      primaryClass={stat.ME},
      url={https://arxiv.org/abs/2309.11914}, 
      howpublished={}
}

@misc{wu2023causalrulelearningenhancing,
      title={Causal Rule Learning: Enhancing the Understanding of Heterogeneous Treatment Effect via Weighted Causal Rules}, 
      author={Ying Wu and Hanzhong Liu and Kai Ren and Xiangyu Chang},
      year={2023},
      eprint={2310.06746},
      archivePrefix={arXiv},
      primaryClass={cs.LG},
      url={https://arxiv.org/abs/2310.06746}, 
      howpublished={}
}

@article{clusteredCATEsurv_Hu2022,
author = {Hu, Liangyuan and Ji, Jiayi and Ennis, Ronald D. and Hogan, Joseph W.},
title = {A flexible approach for causal inference with multiple treatments and clustered survival outcomes},
journal = {Statistics in Medicine},
volume = {41},
number = {25},
pages = {4982-4999},
doi = {https://doi.org/10.1002/sim.9548},
url = {https://onlinelibrary.wiley.com/doi/abs/10.1002/sim.9548},
eprint = {https://onlinelibrary.wiley.com/doi/pdf/10.1002/sim.9548},
year = {2022}
}

@article{tree_subgroup_SIM2023_LuoGuo,
author = {Luo, Yuanhui and Guo, Xinzhou},
title = {Inference on tree-structured subgroups with subgroup size and subgroup effect relationship in clinical trials},
journal = {Statistics in Medicine},
volume = {42},
number = {27},
pages = {5039-5053},
doi = {https://doi.org/10.1002/sim.9900},
url = {https://onlinelibrary.wiley.com/doi/abs/10.1002/sim.9900},
eprint = {https://onlinelibrary.wiley.com/doi/pdf/10.1002/sim.9900},
year = {2023}
}

@article{tutorial_subgroup_SIM2016_Lipkovich,
author = {Lipkovich, Ilya and Dmitrienko, Alex and B. D'Agostino Sr., Ralph},
title = {Tutorial in biostatistics: data-driven subgroup identification and analysis in clinical trials},
journal = {Statistics in Medicine},
volume = {36},
number = {1},
pages = {136-196},
doi = {https://doi.org/10.1002/sim.7064},
url = {https://onlinelibrary.wiley.com/doi/abs/10.1002/sim.7064},
eprint = {https://onlinelibrary.wiley.com/doi/pdf/10.1002/sim.7064},
year = {2017}
}

@article{ChangePoint_subgroup_JASA2017_Fan,
author = {Ailin Fan and Rui Song and Wenbin Lu},
title = {Change-Plane Analysis for Subgroup Detection and Sample Size Calculation},
journal = {Journal of the American Statistical Association},
volume = {112},
number = {518},
pages = {769--778},
year = {2017},
publisher = {ASA Website},
doi = {10.1080/01621459.2016.1166115},

    note ={PMID: 28804182},


URL = { 
    
        https://doi.org/10.1080/01621459.2016.1166115
    
    

},
eprint = { 
    
        https://doi.org/10.1080/01621459.2016.1166115
    
    

}

}

@article{AssesVulnerableSubgroup_JASA2023_Guo,
author = {Xinzhou Guo and Waverly Wei and Molei Liu and Tianxi Cai and Chong Wu and Jingshen Wang},
title = {Assessing the Most Vulnerable Subgroup to Type II Diabetes Associated with Statin Usage: Evidence from Electronic Health Record Data},
journal = {Journal of the American Statistical Association},
volume = {118},
number = {543},
pages = {1488--1499},
year = {2023},
publisher = {ASA Website},
doi = {10.1080/01621459.2022.2157727},

    note ={PMID: 38223220},


URL = { 
    
        https://doi.org/10.1080/01621459.2022.2157727
    
    

},
eprint = { 
    
        https://doi.org/10.1080/01621459.2022.2157727
    
    

}

}

@article{subgroup_recrusive_SIM2011_Lipkovich,
author = {Lipkovich, Ilya and Dmitrienko, Alex and Denne, Jonathan and Enas, Gregory},
title = {Subgroup identification based on differential effect search—A recursive partitioning method for establishing response to treatment in patient subpopulations},
journal = {Statistics in Medicine},
volume = {30},
number = {21},
pages = {2601-2621},
doi = {https://doi.org/10.1002/sim.4289},
url = {https://onlinelibrary.wiley.com/doi/abs/10.1002/sim.4289},
eprint = {https://onlinelibrary.wiley.com/doi/pdf/10.1002/sim.4289},
year = {2011}
}

@article{CATE_simultaneousCI_Biostatistics2011_Cai,
    author = {Cai, Tianxi and Tian, Lu and Wong, Peggy H. and Wei, L. J.},
    title = {Analysis of randomized comparative clinical trial data for personalized treatment selections},
    journal = {Biostatistics},
    volume = {12},
    number = {2},
    pages = {270-282},
    year = {2010},
    month = {09},
    issn = {1465-4644},
    doi = {10.1093/biostatistics/kxq060},
    url = {https://doi.org/10.1093/biostatistics/kxq060},
    eprint = {https://academic.oup.com/biostatistics/article-pdf/12/2/270/18608731/kxq060.pdf},
}

@inproceedings{survRelu_2024_Sun,
author = {Sun, Xiaotong and Qiu, Peijie and Zhang, Shengfan},
title = {SurvReLU: Inherently Interpretable Survival Analysis via Deep ReLU Networks},
year = {2024},
isbn = {9798400704369},
publisher = {Association for Computing Machinery},
address = {New York, NY, USA},
url = {https://doi.org/10.1145/3627673.3679947},
doi = {10.1145/3627673.3679947},
booktitle = {Proceedings of the 33rd ACM International Conference on Information and Knowledge Management},
pages = {4081–4085},
numpages = {5},
location = {Boise, ID, USA},
series = {CIKM '24}
}

%
%

\end{document}